\documentclass[11pt,a4paper,english]{article}
\usepackage{jheppub} 
\usepackage{amsfonts,amsmath,amssymb}
\usepackage[mathscr]{euscript}
\usepackage{wasysym}
\usepackage{stmaryrd}
\usepackage{esint}

\title{Seiberg-Witten geometry of four-dimensional $\mathcal{N}=2$ $\mathrm{SO}-\mathrm{USp}$ quiver gauge theories, I}

\author[a]{Xinyu Zhang}
\affiliation[a]{New High Energy Theory Center and Department of Physics and Astronomy, Rutgers University, \\ Piscataway, New Jersey 08854, USA}

\abstract{
We apply the instanton counting method to study a class of four-dimensional
$\mathcal{N}=2$ supersymmetric quiver gauge theories with alternating
$\mathrm{SO}$ and $\mathrm{USp}$ gauge groups. We compute the partition
function in the $\Omega$-background and express it as functional
integrals over density functions. Applying the saddle point method,
we derive the limit shape equations which determine the dominant instanton
configurations in the flat space limit. The solution to the limit
shape equations gives the Seiberg-Witten geometry of the low energy
effective theory. As an illustrating example, we work out explicitly
the Seiberg-Witten geometry for linear quiver gauge theories. Our
result matches the Seiberg-Witten solution obtained previously using
the method of brane constructions in string theory.
}

\begin{document}
\maketitle


Four-dimensional $\mathcal{N}=2$ supersymmetric gauge theories are
an extremely interesting playground for studying nonperturbative dynamics
of quantum field theories. Following the paradigm of Seiberg and Witten
\cite{Seiberg:1994rs,Seiberg:1994aj}, we can solve exactly the topological
sector of the theory, including the Wilsonian low energy effective
prepotential $\mathcal{F}$ and the correlation functions of gauge-invariant
chiral operators. These quantities receive perturbative corrections
only at one-loop order, while the nonperturbative corrections are
entirely from instantons. The solution is encoded in the data of a
family of complex algebraic curves $\Sigma_{u}$ fibered over the
Coulomb moduli space $\mathcal{B}$, with a meromorphic differential
$\lambda$. 

When the theory admits a microscopic Lagrangian description, a purely
field theoretical algorithm for the derivation of the Seiberg-Witten
solution using the multi-instanton calculus was proposed in \cite{Nekrasov:2002qd},
and no conjectured dualities are assumed. In order to introduce a
supersymmetric regulator of the infinite volume of spacetime and also
to simplify the evaluation of the path integral, the four-dimensional
$\mathcal{N}=2$ supersymmetric gauge theory is formulated in the
$\Omega$-background, which is a particular supergravity background
with two deformation parameters $\varepsilon_{1},\varepsilon_{2}$.
The Poincare symmetry of $\mathbb{R}^{4}$ is broken in a rotationally
covariant way. Applying the technique of equivariant localization,
the partition function $\mathcal{Z}$ in the $\Omega$-background
can be written as a statistical sum over special instanton configurations.
In the flat space limit $\varepsilon_{1},\varepsilon_{2}\to0$, where
the theory in the $\Omega$-background approaches the original flat
space theory, the partition function $\mathcal{Z}$ is dominated by
a particular instanton configuration, the so-called limit shape, with
the instanton charge $\sim\frac{1}{\varepsilon_{1}\varepsilon_{2}}$.
Based on field theoretical arguments \cite{Nekrasov:2002qd,Nekrasov:2003rj},
the Seiberg-Witten prepotential $\mathcal{F}$ of the low energy effective
theory can be extracted from the partition function $\mathcal{Z}$
in the following limit, 
\begin{equation}
\mathcal{F}=-\lim_{\varepsilon_{1},\varepsilon_{2}\to0}\varepsilon_{1}\varepsilon_{2}\log\mathcal{Z}\left(\varepsilon_{1},\varepsilon_{2}\right).
\end{equation}

This approach has hitherto been used to derive the Seiberg-Witten
solution for gauge theories with a simple classical gauge group \cite{Nekrasov:2003rj,Shadchin:2004yx,Shadchin:2005cc},
and $\mathrm{SU}(N)$ quiver gauge theories with hypermultiplets in
the fundamental, adjoint or bifundamental representations \cite{Nekrasov:2012xe}.
However, these are far from all the $\mathcal{N}=2$ supersymmetric
gauge theories for which we are able to compute the partition function
in the $\Omega$-background. It is certainly interesting to cover
all the possible cases and test the validity of this approach.

In this paper, we will be dealing with mass-deformed four-dimensional
$\mathcal{N}=2$ superconformal quiver gauge theories with alternating
$\mathrm{SO}$ and $\mathrm{USp}$ gauge groups. Restricting ourselves
to superconformal theories does not result in a loss of generality,
since asymptotically free theories can always be obtained from superconformal
theories by taking proper scaling limits and decoupling a number of
fundamental hypermultiplets. Generalizing previous computations for
a single $\mathrm{SO}$ or $\mathrm{USp}$ gauge group \cite{Marino:2004cn,Nekrasov:2004vw,Shadchin:2004yx,Fucito:2004gi},
we can compute the partition function in the $\Omega$-background
for $\mathrm{SO}-\mathrm{USp}$ quiver gauge theories. More precisely,
we write the partition function in terms of contour integrals, and
it is not necessary to give the prescription for choosing the correct
poles. One important ingredient in the computation is the treatment
of half-hypermultiplets. Although we cannot compute their contributions
to the partition function directly, we will follow the conjecture
made in \cite{Hollands:2010xa,Hollands:2011zc} that the contribution
of a half-hypermultiplet is given by the square-root of the contribution
of a massless full hypermultiplet composed of a pair of half-hypermultiplets.
Similar to the $\mathrm{SU}$ quiver gauge theories, the limit shape
equations give the gluing conditions of the amplitude functions. The
Seiberg-Witten geometry is finally derived by constructing the functions
invariant under the instanton Weyl group \cite{Nekrasov:2012xe}.
As a representative example, we will describe in great detail the
Seiberg-Witten geometry for linear quiver gauge theories. Our result
matches the Seiberg-Witten solutions obtained previously \cite{Evans:1997hk,Landsteiner:1997vd,Brandhuber:1997cc,Tachikawa:2009rb,Tachikawa:2010vg,Chacaltana:2011ze,Chacaltana:2013oka}. 

The rest of the paper is organized as follows. In Sec. \ref{sec:quiver}
we describe the four-dimensional $\mathcal{N}=2$ $\mathrm{SO}-\mathrm{USp}$
superconformal quiver gauge theories we are dealing with. We introduce
a biparticle quiver diagram to represent the theory. In section \ref{sec:Omega}
we compute explicitly the partition function in the $\Omega$-background.
In the flat space limit, we rewrite the partition function as a functional
integral over density functions. In Sec. \ref{sec:limit shape}, we
apply the saddle point method to determine the special instanton configuration
which dominates the partition function in the flat space limit $\varepsilon_{1},\varepsilon_{2}\to0$.
We solve the limit shape equations by constructing the characters
invariant under the instanton Weyl group. We write down the Seiberg-Witten
curve using the characters. In Sec. \ref{sec:Examples}, we describe
explicitly the Seiberg-Witten geometry for linear $\mathrm{SO}-\mathrm{USp}$
quiver gauge theories. In Sec. \ref{sec:Further} we sketch some possible
further developments of our work. In Appendix \ref{sec:Instanton},
we review the Atiyah-Drinfeld-Hitchin-Manin (ADHM) construction of
instanton moduli space for all classical gauge groups.

\section{$\mathrm{SO}-\mathrm{USp}$ quiver gauge theories \label{sec:quiver}}

A major subtlety of $\mathrm{SO}-\mathrm{USp}$ quiver gauge theories
compared with $\mathrm{SU}$ quiver gauge theories is the appearance
of half-hypermultiplets. Recall that the $\mathcal{N}=2$ hypermultiplet
in the representation $R$ of the gauge group $G$ consists of a pair
of $\mathcal{N}=1$ chiral multiplets, one in the representation $R$
and another in the conjugate representation $R^{*}$ of $G$. If the
representation $R$ is pseudoreal, a single $\mathcal{N}=1$ chiral
superfield forms a consistent $\mathcal{N}=2$ multiplet, the half-hypermultiplet.
From the representation theory of Lie groups $\mathrm{SO}(N)$ and
$\mathrm{USp}(2N)$, we know that the fundamental representation of
$\mathrm{SO}(N)$ is strictly real, while the fundamental representation
of $\mathrm{USp}(2N)$ is pseudo-real. The bifundamental representation
of $\mathrm{SO}(N_{1})\times\mathrm{USp}\left(2N_{2}\right)$, which
is the tensor product of the fundamental representation of $\mathrm{SO}(N_{1})$
and the fundamental representation of $\mathrm{USp}\left(2N_{2}\right)$,
is also pseudoreal. When we couple an $\mathrm{SO}(N)$ vector multiplet
to $N_{f}$ fundamental hypermultiplets, the flavor symmetry group
is $\mathrm{USp}\left(2N_{f}\right)$, and the gauge coupling constant
is marginal when $N_{f}=N-2$. Meanwhile, when we couple an $\mathrm{USp}(2N)$
vector multiplet to $N_{f}$ fundamental half-hypermultiplets, the
flavor symmetry group is $\mathrm{SO}\left(N_{f}\right)$, and the
gauge coupling constant is marginal when $N_{f}=4N+4$. Therefore,
there is a natural way to construct superconformal quiver gauge theories
with alternating $\mathrm{SO}$ and $\mathrm{USp}$ gauge groups.
We certainly cannot avoid half-hypermultiplets in such $\mathrm{SO}-\mathrm{USp}$
quiver gauge theories.

We represent such an $\mathrm{SO}-\mathrm{USp}$ quiver gauge theory
by a bipartite quiver diagram $\gamma$, which consists of vertices
which are colored either black or white and edges connecting vertices
of different colors. The set of vertices is denoted by $V_{\gamma}=V_{\gamma}^{\fullmoon}\cup V_{\gamma}^{\newmoon}$,
where each vertex $i\in V_{\gamma}^{\fullmoon/\newmoon}$ is associated
with a vector multiplet with $\mathrm{SO}/\mathrm{USp}$ gauge group
$G_{i}$. The total gauge group of the quiver gauge theory is
\begin{equation}
G=\prod_{i\in V_{\gamma}}G_{i}=\cdots\times\mathrm{SO}\left(v_{i}=2n_{i}+\mu_{i}\right)\times\mathrm{USp}\left(v_{i+1}-2=2n_{i+1}\right)\times\cdots,
\end{equation}
with $\mu_{i}\in\left\{ 0,1\right\} $. We also define $\mu_{i}=0$
for all $i\in V_{\gamma}^{\newmoon}$. The microscopic gauge coupling
constant $g_{i}$ and the theta angle $\vartheta_{i}$ are combined
into the complexified gauge couplings,
\begin{equation}
\tau_{i}=\frac{\vartheta_{i}}{2\pi}+\frac{4\pi\mathtt{i}}{g_{i}^{2}}.
\end{equation}
We denote the collection of instanton counting parameters by
\begin{equation}
\underline{\mathtt{q}}=\bigcup_{i\in V_{\gamma}}\left\{ \mathtt{q}_{i}=e^{2\pi\mathtt{i}\tau_{i}}\right\} .
\end{equation}

An edge $e=\left\langle i,j\right\rangle $ connecting a vertex $i\in V_{\gamma}^{\fullmoon}$
with a vertex $j\in V_{\gamma}^{\newmoon}$ represents a half-hypermultiplet
in the bifundamental representation of $\mathrm{SO}(v_{i})\times\mathrm{USp}\left(v_{j+1}-2\right)$.
The set of all edges is denoted by $E_{\gamma}$. Unlike the $\mathrm{SU}$
quiver gauge theories, the edges are not oriented. For simplicity,
we assume that there is no edge connecting a vertex to itself. In
particular, in our analysis we disregard the $\mathcal{N}=2^{*}$
theory, which can be treated separately.

We also couple $w_{i}\in\mathbb{Z}_{\ge0}$ fundamental hypermultiplets
to the gauge group $G_{i}$, and additionally $\xi_{i}\in\left\{ 0,1\right\} $
fundamental half-hypermultiplets to the gauge group $\mathrm{USp}\left(2n_{i}\right)$.
The vanishing of the one-loop beta functions of coupling constants
leads to
\begin{eqnarray}
2v_{i} & = & 2w_{i}+\left(4-2\sigma_{i}\right)+\sum_{\left\langle i,j\right\rangle \in E_{\gamma}}v_{j},\quad i\in V_{\gamma}^{\fullmoon},\nonumber \\
2v_{j} & = & 2w_{j}+\xi_{j}+\sum_{\left\langle i,j\right\rangle \in E_{\gamma}}v_{i},\quad j\in V_{\gamma}^{\newmoon},\label{eq:ADEquiver}
\end{eqnarray}
where $\sigma_{i}$ is the number of edges $\left\langle i,j\right\rangle \in E_{\gamma}$
for the fixed $i\in V_{\gamma}^{\fullmoon}$. Since we always make
sure that the number of half-hypermultiplets for each $\mathrm{USp}$
gauge group is even, the theory is safe under Witten\textquoteright s
global anomaly \cite{Witten:1982fp}. The condition (\ref{eq:ADEquiver})
can be solved in a similar way as the $\mathrm{SU}$ quiver gauge
theories. Following the notation in \cite{Nekrasov:2012xe}, we have
the following classification:
\begin{enumerate}
\item Class I theories. The quiver $\gamma$ is the simply laced Dynkin
diagram of the Lie algebra $A_{r}$, $D_{r}$ or $E_{6,7,8}$. For
$A_{r}$-type quivers, we have two possible choices of coloring for
the biparticle diagram, depending on whether the first vertex is $\mathrm{SO}$
or $\mathrm{USp}$ gauge group. For the $D_{r}$-type and $E_{6,7,8}$-type
quivers, we also have two different choices of coloring, depending
on the gauge group at the trivalent vertex.
\item Class II theories. The quiver $\gamma$ is the simply laced affine
Dynkin diagram of the affine Lie algebra $\hat{A}_{r}$, $\hat{D}_{r}$
or $\hat{E}_{6,7,8}$. For $\hat{A}_{r}$-type quivers, we have the
consistency condition which requires that $r$ should be an odd positive
integer. There is no preferred choice of the first vertex, and we
can always fix the first vertex to be a $\mathrm{SO}$ gauge group.
For the $\hat{D}_{r}$-type and the $\hat{E}_{6,7,8}$-type quivers,
depending on the choices of coloring, we again have two sub-types
according to the gauge groups at the trivalent vertices. Notice that
there is no class II{*} theories as in the $\mathrm{SU}$ quiver gauge
theories (except for the $N=2^{*}$ theories which we neglect).
\item Class III theories. There are some extra bizarre theories with non-Dynkin
type quivers. Such theories have to be discussed case by case. See
\cite{Bhardwaj:2013qia} for a complete list.
\end{enumerate}
We consider the low energy effective theory on the Coulomb branch
$\mathcal{B}$ of the moduli space. The coordinates on the Coulomb
branch $\mathcal{B}$ are given by the vacuum expectation values of
the gauge-invariant polynomials of the scalars $\phi_{i}$ in the
vector multiplet,
\begin{equation}
\underline{u}=\bigcup_{i\in V_{\gamma}}\left\{ u_{i,s},s=1,\cdots,n_{i}\right\} .
\end{equation}
In the weakly coupled regime, the vacuum expectation values of $\phi_{i}$
parametrize the Coulomb branch $\mathcal{B}$ locally,
\begin{eqnarray}
\underline{a} & = & \bigcup_{i\in V_{\gamma}^{\fullmoon}}\left\{ a_{i}=\left\langle \phi_{i}\right\rangle =\mathrm{diag}\left\{ a_{i,1},-a_{i,1},\cdots,a_{i,n_{i}},-a_{i,n_{i}},(0)\right\} \right\} \nonumber \\
 &  & \cup\bigcup_{j\in V_{\gamma}^{\newmoon}}\left\{ a_{j}=\left\langle \phi_{j}\right\rangle =\mathrm{diag}\left\{ a_{j,1},\cdots,a_{j,n_{j}},-a_{j,1},\cdots,-a_{j,n_{i}}\right\} \right\} ,
\end{eqnarray}
where $(0)$ is absent for $\mu_{i}=0$. We also turn on generic mass
deformations for the fundamental hypermultiplets. Notice that a single
half-hypermultiplet does not allow a gauge invariant mass term and
must be massless. We collectively denote the set of masses as
\begin{equation}
\underline{m}=\bigcup_{i\in V_{\gamma}}\left\{ m_{i}=\mathrm{diag}\left\{ m_{i,1},\cdots,m_{i,w_{i}}\right\} \right\} .
\end{equation}
It is convenient to encode the Coulomb moduli $a_{i}$ and masses
$m_{i}$ in the characters of two vector space $\mathbf{N}_{i}$ and
$\mathbf{M}_{i}$ assigned for each vertex $i\in V_{\gamma}$,
\begin{eqnarray}
\mathcal{N}_{i} & = & \mathrm{ch}\left(\mathbf{N}_{i}\right)=\sum_{\alpha}\left(e^{\mathtt{i}a_{i,\alpha}}+e^{-\mathtt{i}a_{i,\alpha}}\right)+\mu_{i},\label{eq:Ni}\\
\mathcal{M}_{i} & = & \mathrm{ch}\left(\mathbf{M}_{i}\right)=\sum_{f}e^{\mathtt{i}m_{i,f}}.
\end{eqnarray}

\section{Partition function of quiver gauge theories in the $\Omega$-background
\label{sec:Omega}}

In this section, we compute the partition function of $\mathrm{SO}-\mathrm{USp}$
quiver gauge theories in the $\Omega$-background. The partition function
contains not only the information of the Seiberg-Witten low energy
effective action on $\mathbb{R}^{4}$, but also the low energy effective
couplings of the theory to supergravity background. For the purpose
of this paper, we keep only the relevant information of the partition
function in the flat space limit and rewrite the partition function
as functional integrals over density functions.

It is useful to introduce the following notations. The conversion
operator $\epsilon$ is defined to map characters into weights,
\begin{equation}
\epsilon\left\{ \sum_{i}n_{i}e^{x_{i}}\right\} =\prod_{i}x_{i}^{n_{i}}.
\end{equation}
When the number of terms is infinite, we adopt the regularization
via the analytic continuation,
\begin{equation}
\epsilon\left\{ \sum_{i}n_{i}e^{x_{i}}\right\} =\exp\left(-\left.\frac{d}{ds}\right|_{s=0}\frac{1}{\Gamma(s)}\int_{0}^{\infty}\frac{d\beta}{\beta}\beta^{s}\left(\sum_{i}n_{i}e^{-\beta x_{i}}\right)\right).\label{eq:reg}
\end{equation}
The dual operator $\vee$ is used to flip of sign of the weights
\begin{equation}
\left(\sum_{i}n_{i}e^{x_{i}}\right)^{\vee}=\left(\sum_{i}n_{i}e^{-x_{i}}\right).
\end{equation}
The scaling operator $[p]$ is used to scale the weights,
\begin{equation}
\left(\sum_{i}n_{i}e^{x_{i}}\right)^{[p]}=\left(\sum_{i}n_{i}e^{px_{i}}\right).
\end{equation}
We denote the $\Omega$-deformation parameters as $\varepsilon_{1},\varepsilon_{2}$,
and define
\begin{equation}
\varepsilon=\varepsilon_{1}+\varepsilon_{2},\quad\varepsilon_{\pm}=\frac{\varepsilon_{1}\pm\varepsilon_{2}}{2}.
\end{equation}
We also introduce
\begin{equation}
q_{1}=e^{\mathtt{i}\varepsilon_{1}},\quad q_{2}=e^{\mathtt{i}\varepsilon_{2}},\quad q_{\pm}=e^{\mathtt{i}\varepsilon_{\pm}},\quad\mathcal{P}=\left(q_{1}^{\frac{1}{2}}-q_{1}^{-\frac{1}{2}}\right)\left(q_{2}^{\frac{1}{2}}-q_{2}^{-\frac{1}{2}}\right).
\end{equation}
Notice that the definition of $\mathcal{P}$ is different from the
standard definition in the $\mathrm{SU}$ quiver gauge theories.

\subsection{Instanton partition function}

A four-dimensional $\mathcal{N}=2$ supersymmetric gauge theory in
the $\Omega$-background preserves a supercharge $\mathcal{Q}$, with
$\mathcal{Q}^{2}$ being a sum of the constant gauge transformation
acting on the framing at infinity, the automorphism transformation
of hypermultiplets, and the spacetime rotation. Hence in the twisted
formulation of the theory $\mathcal{Q}$ becomes the equivariant differential,
with the equivariant group being the product of the gauge group $G$,
the flavor group $G_{F}$, and the rotation group $\mathrm{SO}(4)$.
Let $\mathbb{T}$ be the maximal torus of the equivariant group, with
$\left(\underline{a},\underline{m};\varepsilon_{1},\varepsilon_{2}\right)$
being the coordinates on the complexified Lie algebra of $\mathbb{T}$.
It can be shown using the supersymmetric localization principle that
the infinite-dimensional path integral is reduced to finite-dimensional
equivariant integrals over the moduli space of framed instantons,
\begin{equation}
\mathfrak{M}=\left\{ A\in\mathcal{A}|F^{+}=0\right\} /\mathcal{G}_{\infty},
\end{equation}
where $\mathcal{A}$ is the space of connections of principal bundles
over $\mathbb{R}^{4}$, and $\mathcal{G}_{\infty}$ denotes the group
of frame-preserving gauge transformations. For the quiver gauge theories,
$\mathfrak{M}$ is factorized as 
\begin{equation}
\mathfrak{M}=\bigsqcup_{\underline{k}}\mathfrak{M}_{G,\underline{k}}=\bigsqcup_{\underline{k}}\left(\prod_{i\in V_{\gamma}}\mathfrak{M}_{G_{i},k_{i}}\right),
\end{equation}
where we label the instanton charges by
\begin{equation}
\underline{k}=\left\{ k_{i}=\begin{cases}
\kappa_{i}, & i\in V_{\gamma}^{\fullmoon}\\
2\kappa_{i}+\nu_{i}, & i\in V_{\gamma}^{\newmoon}
\end{cases}\right\} \in\mathbb{Z}_{\geq0}^{|\gamma|},
\end{equation}
and $\mathfrak{M}_{G_{i},k_{i}}$ is the moduli spaces of framed $G_{i}$-instantons
with instanton charge $k_{i}$ (see Appendix \ref{sec:Instanton}
for a review). Then the instanton partition function can be written
as 
\begin{equation}
\mathcal{Z}^{\mathrm{inst}}\left(\underline{\mathtt{q}};\underline{a},\underline{m};\varepsilon_{1},\varepsilon_{2}\right)=\sum_{\underline{k}}\underline{\mathtt{q}}^{\underline{k}}\int_{\mathfrak{M}_{G,\underline{k}}}\mathrm{e}_{\mathbb{T}}\left(\mathscr{E}_{\gamma}\right),
\end{equation}
where 
\begin{equation}
\underline{\mathtt{q}}^{\underline{k}}=\prod_{i\in V_{\gamma}^{\fullmoon}}\mathtt{q}_{i}^{\kappa_{i}}\prod_{i\in V_{\gamma}^{\newmoon}}\mathtt{q}_{i}^{\kappa_{i}+\nu_{i}/2},
\end{equation}
and the integration measure $\mathrm{e}_{\mathbb{T}}\left(\mathscr{E}_{\gamma}\right)$
is the $\mathbb{T}$-equivariant Euler class of the matter bundle
$\mathscr{E}_{\gamma}\to\mathfrak{M}_{G,\underline{k}}$ whose fiber
is the space of the virtual zero modes for the Dirac operator in the
instanton background. According to the Atiyah-Singer equivariant index
formula, we have
\begin{equation}
\mathrm{e}_{\mathbb{T}}\left(\mathscr{E}_{\gamma}\right)=\epsilon\left\{ -\int_{\mathbb{C}^{2}}\mathrm{ch}_{\mathbb{T}}\left(\mathscr{E}_{\gamma}\right)\hat{A}\left(\mathbb{C}^{2}\right)\right\} =\epsilon\left\{ -\frac{\iota_{0}^{*}\mathrm{ch}_{\mathbb{T}}\left(\mathscr{E}_{\gamma}\right)}{\mathcal{P}}\right\} ,
\end{equation}
where $\iota_{0}^{*}$ is the pull-back homomorphism induced by the
inclusion $\iota_{0}:0\times\mathfrak{M}_{G,\underline{k}}\to\mathbb{C}^{2}\times\mathfrak{M}_{G,\underline{k}}$.
Applying the Atiyah-Bott equivariant localization formula we can further
reduce the equivariant integration over $\mathfrak{M}_{G,\underline{k}}$
to a discrete sum over the set $\mathfrak{M}_{G,\underline{k}}^{\mathbb{T}}$
of $\mathbb{T}$-fixed points on $\mathfrak{M}_{G,\underline{k}}$,
\begin{equation}
\mathcal{Z}^{\mathrm{inst}}\left(\underline{\mathtt{q}};\underline{a},\underline{m};\varepsilon_{1},\varepsilon_{2}\right)=\sum_{\underline{k}}\underline{\mathtt{q}}^{\underline{k}}\sum_{p\in\mathfrak{M}_{G,\underline{k}}^{\mathbb{T}}}\frac{1}{\mathrm{e}_{\mathbb{T}}\left(T_{p}\mathfrak{M}_{G,\underline{k}}\right)}\epsilon\left\{ -\frac{\iota_{(0,p)}^{*}\mathrm{ch}_{\mathbb{T}}\left(\mathscr{E}_{\gamma}\right)}{\mathcal{P}}\right\} ,
\end{equation}
where $\iota_{(0,p)}^{*}$ is the pull-back homomorphism induced by
the inclusion $\iota_{(0,p)}:0\times p\to\mathbb{C}^{2}\times\mathfrak{M}_{G,\underline{k}}$.
For the $\mathrm{SO}-\mathrm{USp}$ quiver gauge theories, if we denote
$\mathscr{E}_{i}$ to be the $i$th universal bundle over $\mathbb{C}^{2}\times\mathfrak{M}_{G,\underline{k}}$,
whose fiber over an element $A\in\mathfrak{M}_{G_{i},k_{i}}\subset\mathfrak{M}_{G,\underline{k}}$
is the total space of the bundle $E$ with connection $A$, then $\mathscr{E}_{\gamma}$
is given by
\begin{equation}
\mathscr{E}_{\gamma}=\left[\bigoplus_{i\in V_{\gamma}^{\fullmoon}}\mathbf{M}_{i}\otimes\mathscr{E}_{i}\right]\oplus\left[\bigoplus_{i\in V_{\gamma}^{\newmoon}}\mathbf{M}_{i}\otimes\mathscr{E}_{i}\oplus\xi_{i}\mathscr{E}_{i}^{\left(\frac{1}{2}\right)}\right]\oplus\left[\bigoplus_{\left\langle i,j\right\rangle \in E_{\gamma}}\left(\mathscr{E}_{i}\otimes\mathscr{E}_{j}\right)^{\left(\frac{1}{2}\right)}\right],
\end{equation}
where the superscript $\left(\frac{1}{2}\right)$ means half-hypermultiplets. 

The classification of fixed points $\mathfrak{M}_{G,\underline{k}}^{\mathbb{T}}$
for $\mathrm{SO}/\mathrm{USp}$ gauge group is a complicated problem
\cite{Marino:2004cn,Nekrasov:2004vw,Fucito:2004gi,Hollands:2010xa}.
Nevertheless, it is sufficient for us to represent the instanton partition
function as contour integrals,
\begin{eqnarray}
 &  & \mathcal{Z}^{\mathrm{inst}}\left(\underline{\mathtt{q}};\underline{a},\underline{m};\varepsilon_{1},\varepsilon_{2}\right)\nonumber \\
 & = & \sum_{\underline{k}}\underline{\mathtt{q}}^{\underline{k}}\left(\prod_{i\in V_{\gamma}}\frac{1}{\left|W_{i}\right|}\int\prod_{s}\frac{d\phi_{i,s}}{2\pi}\mathcal{Z}_{i}^{\mathrm{inst,vec}}\mathcal{Z}_{i}^{\mathrm{inst,fund}}\right)\left(\prod_{\left\langle i,j\right\rangle \in E_{\gamma}}\mathcal{Z}_{\left\langle i,j\right\rangle }^{\mathrm{inst,bif}}\right),
\end{eqnarray}
where $\left|W_{i}\right|$ is the order of the dual Weyl group of
$G_{i}$, and the factors $\mathcal{Z}_{i}^{\mathrm{inst,vec}}$,
$\mathcal{Z}_{i}^{\mathrm{inst,fund}}$ and $\mathcal{Z}_{\left\langle i,j\right\rangle }^{\mathrm{inst,bif}}$
are contributions to the instanton partition function from the vector
multiplet, the fundamental matter, and the bifundamental half-hypermultiplet.
The variables $\phi_{i,s}$ in the integral are the weights of the
$\mathbb{T}$-action on the space $\mathbb{K}_{i}$, 
\begin{eqnarray}
i\in V_{\gamma}^{\fullmoon} & : & \mathrm{diag}\left\{ e^{\mathtt{i}\phi_{1}},\cdots,e^{\mathtt{i}\phi_{\kappa_{i}}},e^{-\mathtt{i}\phi_{1}},\cdots,e^{-\mathtt{i}\phi_{\kappa_{i}}}\right\} ,\nonumber \\
i\in V_{\gamma}^{\newmoon} & : & \mathrm{diag}\left\{ e^{\mathtt{i}\phi_{1}},e^{-\mathtt{i}\phi_{1}},\cdots,e^{\mathtt{i}\phi_{\kappa_{i}}},e^{-\mathtt{i}\phi_{\kappa_{i}}},(1)\right\} ,
\end{eqnarray}
where $(1)$ is absent for $\nu_{i}=0$. The equivariant character
of the universal bundle $\mathcal{E}_{i}$ evaluated at the origin
is given by 
\begin{equation}
\mathcal{E}_{i}=\iota_{0}\mathrm{ch}_{\mathbb{T}}\left(\mathscr{E}_{i}\right)=\mathcal{N}_{i}-\mathcal{P}\mathcal{K}_{i},\label{eq:chE}
\end{equation}
where $\mathcal{N}_{i}$ is given in (\ref{eq:Ni}), and
\begin{equation}
\mathcal{K}_{i}=\sum_{r=1}^{\kappa_{i}}\left(e^{\mathtt{i}\phi_{i,r}}+e^{-\mathtt{i}\phi_{i,r}}\right)+\nu_{i}.
\end{equation}

We will compute explicitly the factors $\mathcal{Z}_{i}^{\mathrm{inst,vec}}$,
$\mathcal{Z}_{i}^{\mathrm{inst,fund}}$ and $\mathcal{Z}_{\left\langle i,j\right\rangle }^{\mathrm{inst,bif}}$
in the following. We also compute their expansion around the flat
space limit. 

\subsubsection{Vector multiplets}

In order to compute $\mathcal{Z}_{i}^{\mathrm{inst,vec}}$, we use
the basic fact that the character of the tangent space $T\mathfrak{M}_{G,\underline{k}}$
is dual to the index of Dirac operator in the adjoint representation
twisted by the square-root of the canonical bundle. From the representation
theory, the adjoint representation is isomorphic to the rank-two antisymmetric
or symmetric representation for $\mathrm{SO}$ or $\mathrm{USp}$
group, respectively. Therefore, the equivariant character for the
vector multiplet $G_{i},i\in V_{\gamma}^{\fullmoon}$ is 
\begin{eqnarray}
\chi_{i}^{\mathrm{vec}} & = & \frac{q_{+}}{\mathcal{P}}\left(\frac{\mathcal{E}_{i}^{2}-\mathcal{E}_{i}^{[2]}}{2}\right)\nonumber \\
 & = & \frac{q_{+}}{\mathcal{P}}\left(\frac{\mathcal{N}_{i}^{2}-\mathcal{N}_{i}^{[2]}}{2}\right)-q_{+}\mathcal{N}_{i}\mathcal{K}_{i}\nonumber \\
 &  & +\left(1+q\right)\left(\frac{\mathcal{K}_{i}^{2}+\mathcal{K}_{i}^{[2]}}{2}\right)-\left(q_{1}+q_{2}\right)\left(\frac{\mathcal{K}_{i}^{2}-\mathcal{K}_{i}^{[2]}}{2}\right),
\end{eqnarray}
where the first term is the perturbative contribution, and the contribution
to the instanton partition function from the vector multiplet at the
vertex $i\in V_{\gamma}^{\fullmoon}$ is
\begin{eqnarray}
\mathcal{Z}_{i}^{\mathrm{inst,vec}} & = & \epsilon\left\{ -q_{+}\mathcal{N}_{i}\mathcal{K}_{i}+\left(1+q\right)\left(\frac{\mathcal{K}_{i}^{2}+\mathcal{K}_{i}^{[2]}}{2}\right)-\left(q_{1}+q_{2}\right)\left(\frac{\mathcal{K}_{i}^{2}-\mathcal{K}_{i}^{[2]}}{2}\right)\right\} \nonumber \\
 & = & \left(\frac{\varepsilon}{\varepsilon_{1}\varepsilon_{2}}\right)^{k_{i}}\left(\prod_{r=1}^{k_{i}}\frac{4\phi_{i,r}^{2}\left(4\phi_{i,r}^{2}-\varepsilon^{2}\right)}{\mathbb{A}_{i}\left(\phi_{i,r}+\varepsilon_{+}\right)\mathbb{A}_{i}\left(\phi_{i,r}-\varepsilon_{+}\right)}\right)\left(\frac{\Delta_{i}\left(0\right)\Delta_{i}\left(\varepsilon\right)}{\Delta_{i}\left(\varepsilon_{1}\right)\Delta_{i}\left(\varepsilon_{2}\right)}\right),
\end{eqnarray}
where
\begin{eqnarray}
\mathbb{A}_{i}(x) & = & x^{\mu_{i}}\prod_{\alpha}\left(x^{2}-a_{i,\alpha}^{2}\right),\\
\Delta_{i}(x) & = & \prod_{r<s}\left[\left(\phi_{i,r}+\phi_{i,s}\right)^{2}-x^{2}\right]\left[\left(\phi_{i,r}-\phi_{i,s}\right)^{2}-x^{2}\right].
\end{eqnarray}
Similarly, for $i\in V_{\gamma}^{\newmoon}$, we have
\begin{eqnarray}
\chi_{i}^{\mathrm{vec}} & = & \frac{q_{+}}{2\mathcal{P}}\left(\mathcal{E}_{i}^{2}+\mathcal{E}_{i}^{[2]}\right)\nonumber \\
 & = & \frac{q_{+}}{\mathcal{P}}\left(\frac{\mathcal{N}_{i}^{2}+\mathcal{N}_{i}^{[2]}}{2}\right)-q_{+}\mathcal{N}_{i}\mathcal{K}_{i}\nonumber \\
 &  & +\left(1+q\right)\left(\frac{\mathcal{K}_{i}^{2}-\mathcal{K}_{i}^{[2]}}{2}\right)-\left(q_{1}+q_{2}\right)\left(\frac{\mathcal{K}_{i}^{2}+\mathcal{K}_{i}^{[2]}}{2}\right).
\end{eqnarray}
Again, the first term is the perturbative contribution, and the contribution
to the instanton partition function from the vector multiplet at the
vertex $i\in V_{\gamma}^{\newmoon}$ is
\begin{eqnarray}
\mathcal{Z}_{i}^{\mathrm{inst,vec}} & = & \epsilon\left\{ -q_{+}\mathcal{N}_{i}\mathcal{K}_{i}+\left(1+q\right)\left(\frac{\mathcal{K}_{i}^{2}-\mathcal{K}_{i}^{[2]}}{2}\right)-\left(q_{1}+q_{2}\right)\left(\frac{\mathcal{K}_{i}^{2}+\mathcal{K}_{i}^{[2]}}{2}\right)\right\} \nonumber \\
 & = & \left(\frac{\varepsilon}{\varepsilon_{1}\varepsilon_{2}}\right)^{\kappa_{i}}\left[\prod_{r=1}^{\kappa_{i}}\mathbb{A}_{i}\left(\phi_{i,r}+\varepsilon_{+}\right)\mathbb{A}_{i}\left(\phi_{i,r}-\varepsilon_{+}\right)\left(4\phi_{i,r}^{2}-\varepsilon_{1}^{2}\right)\left(4\phi_{i,r}^{2}-\varepsilon_{2}^{2}\right)\right]^{-1}\nonumber \\
 &  & \times\left[\frac{1}{\varepsilon_{1}\varepsilon_{2}\mathbb{A}_{i}\left(\varepsilon_{+}\right)}\prod_{r=1}^{\kappa_{i}}\frac{\phi_{i,r}^{2}\left(\phi_{i,r}^{2}-\varepsilon^{2}\right)}{\left(\phi_{i,r}^{2}-\varepsilon_{1}^{2}\right)\left(\phi_{i,r}^{2}-\varepsilon_{2}^{2}\right)}\right]^{\nu_{i}}\left(\frac{\Delta_{i}\left(0\right)\Delta_{i}\left(\varepsilon\right)}{\Delta_{i}\left(\varepsilon_{1}\right)\Delta_{i}\left(\varepsilon_{2}\right)}\right).
\end{eqnarray}
In the flat space limit, the dominant instanton configuration contributing
to the instanton partition function has the instanton charge of the
order $\sim\frac{1}{\varepsilon_{1}\varepsilon_{2}}$. Therefore,
we should take the limit $\varepsilon_{1},\varepsilon_{2}\to0$, $\kappa_{i}\to\infty$
while keeping $\varepsilon_{1}\varepsilon_{2}\kappa_{i}\sim\mathcal{O}(1)$
fixed. Using the expansion
\begin{equation}
\log\frac{x^{2}\left(x^{2}-\varepsilon^{2}\right)}{\left(x^{2}-\varepsilon_{1}^{2}\right)\left(x^{2}-\varepsilon_{2}^{2}\right)}=\frac{-2\varepsilon_{1}\varepsilon_{2}}{x^{2}}+\mathcal{O}\left(\left(\varepsilon_{1},\varepsilon_{2}\right)^{4}\right),
\end{equation}
we have for $i\in V_{\gamma}^{\fullmoon}$ 
\begin{eqnarray}
F_{i}^{\mathrm{inst,vec}} & = & -\lim_{\varepsilon_{1},\varepsilon_{2}\to0}\varepsilon_{1}\varepsilon_{2}\log\mathcal{Z}_{i}^{\mathrm{inst,vec}}\nonumber \\
 & = & 2\varepsilon_{1}\varepsilon_{2}\sum_{r=1}^{k_{i}}\left[\left(\frac{1}{2}\mu_{i}-1\right)\log\left(\phi_{i,r}^{2}\right)+\sum_{\alpha}\log\left(\phi_{i,r}^{2}-a_{i,\alpha}^{2}\right)\right]\nonumber \\
 &  & +2\left(\varepsilon_{1}\varepsilon_{2}\right)^{2}\sum_{r<s}^{\kappa_{i}}\left[\frac{1}{\left(\phi_{i,r}+\phi_{i,s}\right)^{2}}+\frac{1}{\left(\phi_{i,r}-\phi_{i,s}\right)^{2}}\right],\label{eq:Fiv1}
\end{eqnarray}
and for $i\in V_{\gamma}^{\newmoon}$ 
\begin{eqnarray}
F_{i}^{\mathrm{inst,vec}} & = & -\lim_{\varepsilon_{1},\varepsilon_{2}\to0}\varepsilon_{1}\varepsilon_{2}\log\mathcal{Z}_{i}^{\mathrm{inst,vec}}\nonumber \\
 & = & 2\varepsilon_{1}\varepsilon_{2}\sum_{r=1}^{k_{i}}\left[\log\left(\phi_{i,r}^{2}\right)+\sum_{\alpha}\log\left(\phi_{i,r}^{2}-a_{i,\alpha}^{2}\right)\right]\nonumber \\
 &  & +2\left(\varepsilon_{1}\varepsilon_{2}\right)^{2}\sum_{r<s}^{\kappa_{i}}\left[\frac{1}{\left(\phi_{i,r}+\phi_{i,s}\right)^{2}}+\frac{1}{\left(\phi_{i,r}-\phi_{i,s}\right)^{2}}\right].\label{eq:Fiv2}
\end{eqnarray}
Notice that the $\nu_{i}$-dependent term drops out because it behaves
as 
\begin{equation}
\left(\varepsilon_{1}\varepsilon_{2}\right)^{2}\sum_{r=1}^{k_{i}}\frac{1}{\phi_{i,r}^{2}},
\end{equation}
which vanishes in the flat space limit.

\subsubsection{Fundamental matter}

The equivariant index of a fundamental hypermultiplet is
\begin{equation}
\chi_{i}^{\mathrm{fund,hyper}}=-\frac{\iota_{0}^{*}\mathrm{ch}_{\mathbb{T}}\left(\mathbf{M}_{i}\otimes\mathscr{E}_{i}\right)}{\mathcal{P}}=-\frac{1}{\mathcal{P}}\mathcal{M}_{i}\mathcal{E}_{i}=-\frac{1}{\mathcal{P}}\mathcal{M}_{i}\mathcal{N}_{i}+\mathcal{M}_{i}\mathcal{K}_{i},
\end{equation}
where the first term is the perturbative contribution, and the second
term gives
\begin{equation}
\mathcal{Z}_{i}^{\mathrm{inst,fund,hyper}}=\epsilon\left\{ \mathcal{M}_{i}\mathcal{K}_{i}\right\} =\prod_{f=1}^{w_{i}}\left[m_{i,f}^{\nu_{i}}\prod_{r=1}^{\kappa_{i}}\left(\phi_{i,r}^{2}-m_{i,f}^{2}\right)\right].
\end{equation}
For $i\in V_{\gamma}^{\newmoon}$, we also need to consider the half-hypermultiplet,
which must be massless. We take the contribution of a fundamental
half-hypermultiplet to be the square-root of a massless fundamental
hypermultiplet \cite{Hollands:2010xa},
\begin{equation}
\mathcal{Z}_{i}^{\mathrm{inst,fund,hh}}=\left(\epsilon\left\{ \mathcal{K}_{i}\right\} \right)^{\frac{1}{2}}=\zeta_{i}\prod_{r=1}^{\kappa_{i}}\phi_{i,r},
\end{equation}
where $\zeta_{i}=\pm$. Combining the fundamental hypermultiplet with
possible half-hypermultiplet, we can write the contribution of the
fundamental matter to the instanton partition function as
\begin{equation}
\mathcal{Z}_{i}^{\mathrm{inst,fund}}=\zeta_{i}^{\xi_{i}}\left(\prod_{f=1}^{w_{i}}m_{i,f}^{\nu_{i}}\right)\prod_{r=1}^{\kappa_{i}}\left[\phi_{i,r}^{\xi_{i}}\prod_{f=1}^{w_{i}}\left(\phi_{i,r}^{2}-m_{i,f}^{2}\right)\right].
\end{equation}
In the flat space limit, 
\begin{eqnarray}
F_{i}^{\mathrm{inst,fund}} & = & -\lim_{\varepsilon_{1},\varepsilon_{2}\to0}\varepsilon_{1}\varepsilon_{2}\log\mathcal{Z}_{i}^{\mathrm{inst,fund}}\nonumber \\
 & = & -\varepsilon_{1}\varepsilon_{2}\sum_{r=1}^{\kappa_{i}}\left[\frac{\varphi_{i}}{2}\log\left(\phi_{i,r}^{2}\right)+\sum_{f=1}^{w_{i}}\log\left(\phi_{i,r}^{2}-m_{i,f}^{2}\right)\right],\label{eq:Fif}
\end{eqnarray}
where the dependence on the overall sign $\zeta_{i}$ disappears.

\subsubsection{Bifundamental half-hypermultiplet}

The equivariant index of a massless bifundamental hypermultiplet composed
of a pair of bifundamental half-hypermultiplets is given by
\begin{eqnarray}
\chi_{\left\langle i,j\right\rangle }^{\mathrm{bif,hyper}} & = & -\frac{1}{\mathcal{P}}\left(\mathcal{N}_{i}-\mathcal{P}\mathcal{K}_{i}\right)\left(\mathcal{N}_{j}-\mathcal{P}\mathcal{K}_{j}\right)\nonumber \\
 & = & -\frac{1}{\mathcal{P}}\mathcal{N}_{i}\mathcal{N}_{j}+\mathcal{N}_{i}\mathcal{K}_{j}+\mathcal{N}_{j}\mathcal{K}_{i}-\mathcal{P}\mathcal{K}_{i}\mathcal{K}_{j},
\end{eqnarray}
where the first term is the perturbative contribution, and the instanton
contribution given by the remaining terms is a complete square,
\begin{eqnarray}
\mathcal{Z}_{\left\langle i,j\right\rangle }^{\mathrm{inst,bif,hyper}} & = & \left[\left(\prod_{\alpha}a_{i,\alpha}\right)\left(\prod_{r=1}^{\kappa_{i}}\frac{\phi_{i,r}^{2}-\varepsilon_{-}^{2}}{\phi_{i,r}^{2}-\varepsilon_{+}^{2}}\right)\right]^{2\nu_{j}}\nonumber \\
 &  & \times\left[\prod_{r=1}^{\kappa_{i}}\mathbb{A}_{j}\left(\phi_{i,r}\right)\right]^{2}\left[\prod_{s=1}^{\kappa_{j}}\mathbb{A}_{i}\left(\phi_{j,s}\right)\right]^{2}\left(\frac{\Delta_{i,j}\left(\varepsilon_{-}\right)}{\Delta_{i,j}\left(\varepsilon_{+}\right)}\right)^{2},
\end{eqnarray}
where 
\begin{equation}
\Delta_{i,j}(x)=\prod_{r=1}^{\kappa_{i}}\prod_{s=1}^{\kappa_{j}}\left[\left(\phi_{i,r}+\phi_{j,s}\right)^{2}-x^{2}\right]\left[\left(\phi_{i,r}-\phi_{j,s}\right)^{2}-x^{2}\right].
\end{equation}
We identify the contribution to the instanton partition function from
the bifundamental half-hypermultiplet as \cite{Hollands:2010xa}
\begin{eqnarray}
\mathcal{Z}_{\left\langle i,j\right\rangle }^{\mathrm{inst,bif}} & = & \zeta_{\left\langle i,j\right\rangle }\left[\left(\prod_{\alpha}a_{i,\alpha}\right)\left(\prod_{r=1}^{\kappa_{i}}\frac{\phi_{i,r}^{2}-\varepsilon_{-}^{2}}{\phi_{i,r}^{2}-\varepsilon_{+}^{2}}\right)\right]^{\nu_{j}}\nonumber \\
 &  & \times\left[\prod_{r=1}^{\kappa_{i}}\mathbb{A}_{j}\left(\phi_{i,r}\right)\right]\left[\prod_{s=1}^{\kappa_{j}}\mathbb{A}_{i}\left(\phi_{j,s}\right)\right]\left(\frac{\Delta_{i,j}\left(\varepsilon_{-}\right)}{\Delta_{i,j}\left(\varepsilon_{+}\right)}\right),
\end{eqnarray}
with the overall sign $\zeta_{\left\langle i,j\right\rangle }=\pm$.
Using the expansion
\begin{equation}
\log\frac{x^{2}-\varepsilon_{-}^{2}}{x^{2}-\varepsilon_{+}^{2}}=\frac{\varepsilon_{1}\varepsilon_{2}}{x^{2}}+\mathcal{O}\left(\left(\varepsilon_{1},\varepsilon_{2}\right)^{4}\right),
\end{equation}
we can compute the flat space limit,
\begin{eqnarray}
F_{\left\langle i,j\right\rangle }^{\mathrm{inst,bif}} & = & -\lim_{\varepsilon_{1},\varepsilon_{2}\to0}\varepsilon_{1}\varepsilon_{2}\log\mathcal{Z}_{\left\langle i,j\right\rangle }^{\mathrm{inst,bif}}\nonumber \\
 & = & -\varepsilon_{1}\varepsilon_{2}\sum_{r=1}^{\kappa_{i}}\sum_{\alpha}\log\left(\phi_{i,r}^{2}-a_{j,\alpha}^{2}\right)\nonumber \\
 &  & -\varepsilon_{1}\varepsilon_{2}\sum_{s=1}^{\kappa_{j}}\left[\frac{\mu_{i}}{2}\log\left(\phi_{j,s}^{2}\right)+\sum_{\alpha}\log\left(\phi_{j,s}^{2}-a_{i,\alpha}^{2}\right)\right]\nonumber \\
 &  & -\left(\varepsilon_{1}\varepsilon_{2}\right)^{2}\sum_{r=1}^{\kappa_{i}}\sum_{s=1}^{\kappa_{j}}\left[\frac{1}{\left(\phi_{i,r}+\phi_{j,s}\right)^{2}}+\frac{1}{\left(\phi_{i,r}-\phi_{j,s}\right)^{2}}\right].\label{eq:Fib}
\end{eqnarray}
We see that the $\nu_{j}$-dependent term and the overall sign $\zeta_{\left\langle i,j\right\rangle }$
drop out again.

\subsection{Full partition function}

After deriving the instanton partition function, we would like to
combine it with the classical and the perturbative contributions to
form the full partition function,
\begin{eqnarray}
 &  & \mathcal{Z}\left(\underline{\mathtt{q}};\underline{a},\underline{m};\varepsilon_{1},\varepsilon_{2}\right)\nonumber \\
 & = & \mathcal{Z}^{\mathrm{cl}}\mathcal{Z}^{\mathrm{pert}}\mathcal{Z}^{\mathrm{inst}}\nonumber \\
 & = & \mathcal{Z}^{\mathrm{cl}}\sum_{\underline{k}}\underline{\mathtt{q}}^{\underline{k}}\left(\prod_{i\in V_{\gamma}}\frac{1}{\left|W_{i}\right|}\int\prod_{s}\frac{d\phi_{i,s}}{2\pi}\mathcal{Z}_{i}^{\mathrm{pert,vec}}\mathcal{Z}_{i}^{\mathrm{pert,fund}}\mathcal{Z}_{i}^{\mathrm{inst,vec}}\mathcal{Z}_{i}^{\mathrm{inst,fund}}\right)\nonumber \\
 &  & \times\left(\prod_{\left\langle i,j\right\rangle \in E_{\gamma}}\mathcal{Z}_{\left\langle i,j\right\rangle }^{\mathrm{pert,bif}}\mathcal{Z}_{\left\langle i,j\right\rangle }^{\mathrm{inst,bif}}\right).
\end{eqnarray}

The classical partition function is simply given by
\begin{eqnarray*}
\mathcal{Z}^{\mathrm{cl}}\left(\underline{\mathtt{q}};\underline{a};\varepsilon_{1},\varepsilon_{2}\right) & = & \prod_{i\in V_{\gamma}}\mathtt{q}_{i}^{-\frac{1}{2\varepsilon_{1}\varepsilon_{2}}\sum_{\alpha}a_{i,\alpha}^{2}}
\end{eqnarray*}
whose flat space limit is 
\begin{equation}
F^{\mathrm{cl}}=-\lim_{\varepsilon_{1},\varepsilon_{2}\to0}\varepsilon_{1}\varepsilon_{2}\log\mathcal{Z}^{\mathrm{cl}}=\frac{1}{2}\sum_{i\in V_{\gamma}}\log\left(\mathtt{q}_{i}\right)\sum_{\alpha}a_{i,\alpha}^{2}.\label{eq:Fc}
\end{equation}

The perturbative contribution to the partition function in the $\Omega$-background
is one-loop exact. In fact, we have already obtained them as the byproduct
of our derivation of the instanton partition function,
\begin{eqnarray}
\mathcal{Z}_{i\in V_{\gamma}^{\fullmoon}}^{\mathrm{pert,vec}} & = & \epsilon\left\{ \frac{q_{+}}{\mathcal{P}}\left(\frac{\mathcal{N}_{i}^{2}-\mathcal{N}_{i}^{[2]}}{2}\right)\right\} ,\\
\mathcal{Z}_{i\in V_{\gamma}^{\newmoon}}^{\mathrm{pert,vec}} & = & \epsilon\left\{ \frac{q_{+}}{\mathcal{P}}\left(\frac{\mathcal{N}_{i}^{2}+\mathcal{N}_{i}^{[2]}}{2}\right)\right\} ,\\
\mathcal{Z}_{i}^{\mathrm{pert,fund}} & = & \epsilon\left\{ -\frac{1}{\mathcal{P}}\left(\mathcal{M}_{i}+\frac{1}{2}\varphi_{i}\right)\mathcal{N}_{i}\right\} ,\\
\mathcal{Z}_{\left\langle i,j\right\rangle }^{\mathrm{pert,bif}} & = & \epsilon\left\{ -\frac{1}{2\mathcal{P}}\mathcal{N}_{i}\mathcal{N}_{j}\right\} .
\end{eqnarray}
We set the cutoff energy scale $\Lambda_{\mathrm{UV}}=1$. Using the
regularization (\ref{eq:reg}), we can write the perturbative contributions
in terms of Barnes' double Gamma function $\Gamma_{2}\left(\left.x\right|\varepsilon_{1},\varepsilon_{2}\right)$.
Apply the relation
\begin{equation}
-\lim_{\varepsilon_{1},\varepsilon_{2}\to0}\varepsilon_{1}\varepsilon_{2}\log\Gamma_{2}\left(\left.x\right|\varepsilon_{1},\varepsilon_{2}\right)=\frac{x^{2}}{4}\log\left(x^{2}\right)-\frac{3x^{2}}{4}=\mathscr{K}(x),
\end{equation}
we get the perturbative contributions in the flat space limit,
\begin{eqnarray}
F_{i\in V_{\gamma}^{\fullmoon}}^{\mathrm{pert,vec}} & = & -\lim_{\varepsilon_{1},\varepsilon_{2}\to0}\varepsilon_{1}\varepsilon_{2}\log\mathcal{Z}_{i}^{\mathrm{pert,vec}}\nonumber \\
 & = & -\sum_{\alpha<\beta}\mathscr{K}\left(a_{\alpha}+a_{\beta}\right)-\sum_{\alpha<\beta}\mathscr{K}\left(a_{\alpha}-a_{\beta}\right)+2\mu_{i}\sum_{\alpha}\mathscr{K}\left(a_{\alpha}\right),
\end{eqnarray}
\begin{eqnarray}
F_{i\in V_{\gamma}^{\newmoon}}^{\mathrm{pert,vec}} & = & -\lim_{\varepsilon_{1},\varepsilon_{2}\to0}\varepsilon_{1}\varepsilon_{2}\log\mathcal{Z}_{i}^{\mathrm{pert,vec}}\nonumber \\
 & = & -\sum_{\alpha\leq\beta}\mathscr{K}\left(a_{\alpha}+a_{\beta}\right)-\sum_{\alpha<\beta}\mathscr{K}\left(a_{\alpha}-a_{\beta}\right),
\end{eqnarray}
\begin{eqnarray}
F_{i}^{\mathrm{pert,fund}} & = & -\lim_{\varepsilon_{1},\varepsilon_{2}\to0}\varepsilon_{1}\varepsilon_{2}\log\mathcal{Z}_{i}^{\mathrm{pert,fund}}\nonumber \\
 & = & \sum_{\alpha}\sum_{f=1}^{w_{i}}\left[\mathscr{K}\left(a_{i,\alpha}+m_{i,f}\right)+\mathscr{K}\left(a_{i,\alpha}-m_{i,f}\right)\right]\nonumber \\
 &  & +\xi_{i}\sum_{\alpha}\mathscr{K}\left(a_{i,\alpha}\right)+\mu_{i}\sum_{f=1}^{w_{i}}\mathscr{K}\left(m_{i,f}\right),
\end{eqnarray}
\begin{eqnarray}
F_{\left\langle i,j\right\rangle }^{\mathrm{pert,bif}} & = & -\lim_{\varepsilon_{1},\varepsilon_{2}\to0}\varepsilon_{1}\varepsilon_{2}\log\mathcal{Z}_{\left\langle i,j\right\rangle }^{\mathrm{pert,bif}}\nonumber \\
 & = & \sum_{\alpha,\beta}\left[\mathscr{K}\left(a_{i,\alpha}+a_{j,\beta}\right)+\mathscr{K}\left(a_{i,\alpha}-a_{j,\beta}\right)\right]+\mu_{i}\sum_{\beta}\mathscr{K}\left(a_{j,\beta}\right).
\end{eqnarray}

Therefore, the full partition function in the flat space limit can
be written as
\begin{equation}
\mathcal{Z}\left(\underline{\mathtt{q}};\underline{a},\underline{m};\varepsilon_{1},\varepsilon_{2}\right)=\sum_{\underline{k}}\underline{\mathtt{q}}^{\underline{k}}\int\prod_{i\in V_{\gamma}}\prod_{s}\frac{d\phi_{i,s}}{2\pi}\exp\left(-\frac{F}{\varepsilon_{1}\varepsilon_{2}}\right),\label{eq:Zfull}
\end{equation}
where 
\begin{eqnarray}
F & = & F^{\mathrm{cl}}+\sum_{i\in V_{\gamma}}\left(F_{i}^{\mathrm{pert,vec}}+F_{i}^{\mathrm{pert,fund}}+F_{i}^{\mathrm{inst,vec}}+F_{i}^{\mathrm{inst,fund}}\right)\nonumber \\
 &  & +\sum_{\left\langle i,j\right\rangle \in E_{\gamma}}\left(F_{\left\langle i,j\right\rangle }^{\mathrm{pert,bif}}+F_{\left\langle i,j\right\rangle }^{\mathrm{inst,bif}}\right)+\mathcal{O}\left(\varepsilon_{1},\varepsilon_{2}\right).
\end{eqnarray}

\subsection{The functional integrals over density functions}

It is useful to rewrite $F$ in terms of functional integrals of density
functions. We introduce the instanton density functions
\begin{equation}
\varrho_{i}(z)=\varepsilon_{1}\varepsilon_{2}\sum_{s=1}^{\kappa_{i}}\left[\delta\left(z-\phi_{i,s}\right)+\delta\left(z+\phi_{i,s}\right)\right],
\end{equation}
with the normalization ensuring the finiteness in the flat space limit.
They are even functions,
\begin{equation}
\varrho_{i}(z)=\varrho_{i}(-z).
\end{equation}
Using the standard rule 
\begin{equation}
\varepsilon_{1}\varepsilon_{2}\sum_{r=1}^{\kappa_{i}}\left[f\left(\phi_{i,r}\right)+f\left(-\phi_{i,r}\right)\right]\to\int dz\varrho_{i}(z)f(z),
\end{equation}
we can rewrite $F_{i}^{\mathrm{inst,vec}}$, $F_{i}^{\mathrm{inst,fund}}$,
and $F_{\left\langle i,j\right\rangle }^{\mathrm{inst,bif}}$ in terms
of $\varrho_{i}(z)$ as
\begin{eqnarray}
F_{i\in V_{\gamma}^{\fullmoon}}^{\mathrm{inst,vec}} & = & 2\int dz\varrho_{i}(z)\left[\left(\frac{1}{2}\mu_{i}-1\right)\log(z)+\sum_{\alpha}\log\left(z-a_{i,\alpha}\right)\right]\nonumber \\
 &  & +\frac{1}{2}\fint dzdz^{\prime}\frac{\varrho_{i}(z)\varrho_{i}(z^{\prime})}{(z-z^{\prime})^{2}}\nonumber \\
 & = & 2\int dz\varrho_{i}^{\prime\prime}(z)\left[\left(\frac{1}{2}\mu_{i}-1\right)\mathscr{K}(z)+\sum_{\alpha}\mathscr{K}\left(z-a_{i,\alpha}\right)\right]\nonumber \\
 &  & -\frac{1}{2}\fint dzdz^{\prime}\varrho_{i}^{\prime\prime}(z)\varrho_{i}^{\prime\prime}(z^{\prime})\mathscr{K}\left(z-z^{\prime}\right),
\end{eqnarray}
\begin{eqnarray}
F_{i\in V_{\gamma}^{\newmoon}}^{\mathrm{inst,vec}} & = & 2\int dz\varrho_{i}(z)\left[\log(z)+\sum_{\alpha}\log\left(z-a_{i,\alpha}\right)\right]\nonumber \\
 &  & +\frac{1}{2}\fint dzdz^{\prime}\frac{\varrho_{i}(z)\varrho_{i}(z^{\prime})}{(z-z^{\prime})^{2}}\nonumber \\
 & = & 2\int dz\varrho_{i}^{\prime\prime}(z)\left[\mathscr{K}(z)+\sum_{\alpha}\mathscr{K}\left(z-a_{i,\alpha}\right)\right]\nonumber \\
 &  & -\frac{1}{2}\fint dzdz^{\prime}\varrho_{i}^{\prime\prime}(z)\varrho_{i}^{\prime\prime}(z^{\prime})\mathscr{K}\left(z-z^{\prime}\right),
\end{eqnarray}
\begin{eqnarray}
F_{i}^{\mathrm{inst,fund}} & = & -\int dz\varrho_{i}(z)\left[\frac{\xi_{i}}{2}\log(z)+\sum_{f=1}^{w_{i}}\log\left(z-m_{i,f}\right)\right]\nonumber \\
 & = & -\int dz\varrho_{i}^{\prime\prime}(z)\left[\frac{\xi_{i}}{2}\mathscr{K}(z)+\sum_{f=1}^{w_{i}}\mathscr{K}\left(z-m_{i,f}\right)\right],
\end{eqnarray}
\begin{eqnarray}
F_{\left\langle i,j\right\rangle }^{\mathrm{inst,bif}} & = & -\int dz\varrho_{i}(z)\sum_{\alpha}\log\left(z-a_{j,\alpha}\right)\nonumber \\
 &  & -\int dz\varrho_{j}(z)\left[\frac{\mu_{i}}{2}\log(z)+\sum_{\alpha}\log\left(z-a_{i,\alpha}\right)\right]\nonumber \\
 &  & -\frac{1}{2}\fint dzdz^{\prime}\frac{\varrho_{i}(z)\varrho_{j}(z^{\prime})}{(z-z^{\prime})^{2}}\nonumber \\
 & = & -\int dz\varrho_{i}^{\prime\prime}(z)\sum_{\alpha}\mathscr{K}\left(z-a_{j,\alpha}\right)\nonumber \\
 &  & -\int dz\varrho_{j}^{\prime\prime}(z)\left[\frac{\mu_{i}}{2}\mathscr{K}(z)+\sum_{\alpha}\mathscr{K}\left(z-a_{i,\alpha}\right)\right]\nonumber \\
 &  & +\frac{1}{2}\fint dzdz^{\prime}\varrho_{i}^{\prime\prime}(z)\varrho_{j}^{\prime\prime}(z^{\prime})\mathscr{K}\left(z-z^{\prime}\right),
\end{eqnarray}
where $\fint$ denotes the principal value of the improper integral. 

In order to combine the instanton contribution with the perturbative
contribution, we introduce the full density functions
\begin{eqnarray}
\rho_{i}(z) & = & \begin{cases}
\mu_{i}\delta(z)+\sum_{\alpha}\left[\delta\left(z-a_{i,\alpha}\right)+\delta\left(z+a_{i,\alpha}\right)\right]-\varrho^{\prime\prime}(z), & i\in V_{\gamma}^{\fullmoon}\\
2\delta\left(z\right)+\sum_{\alpha}\left[\delta\left(z-a_{i,\alpha}\right)+\delta\left(z+a_{i,\alpha}\right)\right]-\varrho^{\prime\prime}(z), & i\in V_{\gamma}^{\newmoon}
\end{cases},\label{eq:density}
\end{eqnarray}
so that
\begin{eqnarray}
F_{i}^{\mathrm{vec}} & = & F_{i}^{\mathrm{pert,vec}}+F_{i}^{\mathrm{inst,vec}}\nonumber \\
 & = & \begin{cases}
-\frac{1}{2}\fint dzdz^{\prime}\rho_{i}(z)\rho_{i}(z^{\prime})\mathscr{K}\left(z-z^{\prime}\right)+2\int dz\rho_{i}(z)\mathscr{K}(z), & i\in V_{\gamma}^{\fullmoon}\\
-\frac{1}{2}\fint dzdz^{\prime}\rho_{i}(z)\rho_{i}(z^{\prime})\mathscr{K}\left(z-z^{\prime}\right), & i\in V_{\gamma}^{\newmoon}
\end{cases},\label{eq:Fv}
\end{eqnarray}
\begin{eqnarray}
F_{i}^{\mathrm{fund}} & = & F_{i}^{\mathrm{pert,fund}}+F_{i}^{\mathrm{inst,fund}}\nonumber \\
 & = & \sum_{f=1}^{w_{i}}\int dz\rho_{i}(z)\mathscr{K}\left(z-m_{i,f}\right)+\frac{\xi_{i}}{2}\int dz\rho_{i}(z)\mathscr{K}\left(z\right),\label{eq:Ff}
\end{eqnarray}
\begin{eqnarray}
F_{\left\langle i,j\right\rangle }^{\mathrm{bif}} & = & F_{\left\langle i,j\right\rangle }^{\mathrm{pert,bif}}+F_{\left\langle i,j\right\rangle }^{\mathrm{inst,bif}}\nonumber \\
 & = & \frac{1}{2}\fint dzdz^{\prime}\rho_{i}(z)\rho_{j}(z^{\prime})\mathscr{K}\left(z-z^{\prime}\right)-\int dz\rho_{i}(z)\mathscr{K}(z),\label{eq:Fb}
\end{eqnarray}
where we have used the fact that $\mathscr{K}(0)=0$. Clearly the
expressions (\ref{eq:Fv})(\ref{eq:Ff})(\ref{eq:Fb}) are much simpler
than the perturbative and instanton contributions separately, and
the explicit dependence of the Coulomb moduli $a_{i,\alpha}$ disappears. 

For the classical contribution and the factor $\underline{\mathtt{q}}^{\underline{k}}$,
we can evaluate that
\begin{equation}
\int z^{2}\rho_{i}(z)dz=2\sum_{\alpha}a_{i,\alpha}^{2}-4\varepsilon_{1}\varepsilon_{2}\kappa_{i},
\end{equation}
which leads to 
\begin{eqnarray}
-\varepsilon_{1}\varepsilon_{2}\log\left(\mathcal{Z}^{\mathrm{cl}}\underline{\mathtt{q}}^{\underline{k}}\right) & = & \sum_{i\in V_{\gamma}}\log\left(\mathtt{q}_{i}\right)\left(\frac{1}{2}\sum_{\alpha}a_{i,\alpha}^{2}-\varepsilon_{1}\varepsilon_{2}\left(\kappa_{i}+\frac{1}{2}\nu_{i}\right)\right)\nonumber \\
 & = & \frac{1}{4}\sum_{i\in V_{\gamma}}\log\left(\mathtt{q}_{i}\right)\int z^{2}\rho_{i}(z)dz+\mathcal{O}\left(\varepsilon_{1},\varepsilon_{2}\right).
\end{eqnarray}

Therefore, we can rewrite the full partition function in terms of
functional integrals over $\rho=\left\{ \rho_{i}\right\} _{i\in V_{\gamma}}$,

\begin{equation}
\mathcal{Z}=\int\prod_{i\in V_{\gamma}}d\rho_{i}\exp\left(-\frac{F[\rho]}{\varepsilon_{1}\varepsilon_{2}}+\mathcal{O}\left(\varepsilon_{1},\varepsilon_{2}\right)\right),
\end{equation}
where $F[\rho]$ is given by
\begin{eqnarray}
F[\rho] & = & -\frac{1}{2}\sum_{i\in V_{\gamma}}\fint dzdz^{\prime}\rho_{i}(z)\rho_{i}(z^{\prime})\mathscr{K}\left(z-z^{\prime}\right)\nonumber \\
 &  & +\sum_{i\in V_{\gamma}^{\fullmoon}}\int dz\rho_{i}(z)\left[2\mathscr{K}(z)+\frac{1}{4}\log\left(\mathtt{q}_{i}\right)z^{2}+\sum_{f=1}^{w_{i}}\mathscr{K}\left(z-m_{i,f}\right)\right]\nonumber \\
 &  & +\sum_{i\in V_{\gamma}^{\newmoon}}\int dz\rho_{i}(z)\left[\frac{1}{4}\log\left(\mathtt{q}_{i}\right)z^{2}+\sum_{f=1}^{w_{i}}\mathscr{K}\left(z-m_{i,f}\right)+\frac{\xi_{i}}{2}\mathscr{K}(z)\right]\nonumber \\
 &  & +\sum_{\left\langle i,j\right\rangle \in E_{\gamma}}\left[\frac{1}{2}\fint dzdz^{\prime}\rho_{i}(z)\rho_{j}(z^{\prime})\mathscr{K}\left(z-z^{\prime}\right)-\int dz\rho_{i}(z)\mathscr{K}(z)\right].
\end{eqnarray}

\section{The limit shape equations \label{sec:limit shape}}

Now we are ready to perform the saddle-point evaluation following
the approach in \cite{Nekrasov:2003rj,Nekrasov:2004vw,Shadchin:2004yx,Shadchin:2005cc,Nekrasov:2012xe}
to determine the limit shape of the instanton configuration which
dominates the partition function (\ref{eq:Zfull}) in the flat space
limit $\varepsilon_{1},\varepsilon_{2}\to0$. 

\subsection{Saddle point analysis}

In the limit $\varepsilon_{1},\varepsilon_{2}\to0$, the distribution
$\rho_{i}(z)$ becomes a function with compact support $\mathscr{C}_{i}$.
In an appropriate domain of the parameter space, $\mathscr{C}_{i}$
for different $i$ are widely separated, and each $\mathscr{C}_{i}$
is a union of disjoint intervals along the real axis,
\begin{equation}
\mathscr{C}_{i}=\bigcup_{\ell}\mathscr{I}_{i,\ell}=\bigcup_{\ell}\left[a_{i,\ell}^{-},a_{i,\ell}^{+}\right],\quad\cdots<a_{i,\ell}^{-}<a_{i,\ell}^{+}<a_{i,\ell+1}^{-}<a_{i,\ell+1}^{+}<\cdots.
\end{equation}
Here
\begin{eqnarray*}
\ell & = & \begin{cases}
\pm1,\pm2,\cdots,\pm n_{i}, & i\in V_{\gamma}^{\fullmoon}\\
0,\pm1,\pm2,\cdots,\pm n_{i}, & i\in V_{\gamma}^{\newmoon}
\end{cases},
\end{eqnarray*}
with $\pm a_{i,\alpha}\in\mathscr{I}_{i,\pm\alpha}$ and $0\in\mathscr{I}_{i,0}$.
The function $\rho_{i}(z)$ is normalized according to
\begin{equation}
\int_{\mathscr{I}_{i,\ell}}\rho_{i}(z)dz=\begin{cases}
2, & i\in V_{\gamma}^{\newmoon},\ell=0\\
1, & \mathrm{otherwise}
\end{cases}.
\end{equation}
The Coulomb moduli enter the variational problem via the additional
constraints 
\begin{equation}
\int_{\mathscr{I}_{i,\pm\alpha}}z\rho_{i}(z)dz=\pm a_{i,\alpha}.\label{eq:inta}
\end{equation}
After incorporating the constraints via Lagrangian multipliers $b_{i,\alpha}$
and $a_{i,\alpha}^{D}$, our task is to find the limit shape $\rho_{\star}$
which extremizes the following effective free energy,
\begin{equation}
F^{\mathrm{eff}}[\rho]=F[\rho]+\sum_{i\in V_{\gamma}}\sum_{\ell}\left[b_{i,\ell}\left(1-\int_{\mathscr{I}_{i,\ell}}\rho_{i}(z)dz\right)+a_{i,\ell}^{D}\left(a_{i,\ell}-\int_{\mathscr{I}_{i,\ell}}z\rho_{i}(z)dz\right)\right],
\end{equation}
where $a_{i,\alpha}^{D}$ is the dual special coordinate of $a_{i,\alpha}$,
and the low energy effective prepotential is given in terms of $\rho_{\star}$
as 
\begin{equation}
\mathcal{F}=F^{\mathrm{eff}}[\rho_{\star}].
\end{equation}

For any $i\in V_{\gamma}^{\fullmoon}$ and $x\in\mathscr{I}_{i,\ell}$,
the variation of $F^{\mathrm{eff}}[\rho]$ with respect to $\rho_{i}(x)$
leads to the following linear integral equation,
\begin{eqnarray}
0 & = & -\int dz\rho_{i}(z)\mathscr{K}\left(z-x\right)+2\mathscr{K}(x)+\frac{1}{4}\log\left(\mathtt{q}_{i}\right)x^{2}\nonumber \\
 &  & +\frac{1}{2}\sum_{f=1}^{w_{i}}\left[\mathscr{K}\left(x-m_{i,f}\right)+\mathscr{K}\left(x+m_{i,f}\right)\right]\nonumber \\
 &  & +\sum_{\left\langle i,j\right\rangle \in E_{\gamma}}\left[\frac{1}{2}\int dz\rho_{j}(z)\mathscr{K}\left(x-z\right)-\mathscr{K}(x)\right]-b_{i,\ell}-xa_{i,\ell}^{D}.
\end{eqnarray}
Keep in mind that we must preserve the symmetry $\rho_{i}(x)=\rho_{i}(-x)$
in the variation. Similarly, for any $j\in V_{\gamma}^{\newmoon}$
and $x\in\mathscr{I}_{j,\ell}$, we have
\begin{eqnarray}
0 & = & -\int dz\rho_{j}(z)\mathscr{K}\left(z-x\right)+\frac{1}{4}\log\left(\mathtt{q}_{j}\right)x^{2}\nonumber \\
 &  & +\frac{1}{2}\sum_{f=1}^{w_{j}}\left[\mathscr{K}\left(x-m_{j,f}\right)+\mathscr{K}\left(x+m_{j,f}\right)\right]+\frac{\xi_{j}}{2}\mathscr{K}(x)\nonumber \\
 &  & +\frac{1}{2}\sum_{\left\langle i,j\right\rangle \in E_{\gamma}}\int dz\rho_{i}(z)\mathscr{K}\left(z-x\right)-b_{j,\ell}-xa_{j,\ell}^{D}.
\end{eqnarray}
Taking the second derivative with respect to $x$, we obtain the limit
shape equations
\begin{eqnarray}
0 & = & -\fint dz\rho_{i}(z)\log\left(x-z\right)+2\log x+\frac{1}{2}\log\left(\mathtt{q}_{i}\right)+\frac{1}{2}\sum_{f=1}^{w_{i}}\log\left(x^{2}-m_{i,f}^{2}\right)\nonumber \\
 &  & +\sum_{\left\langle i,j\right\rangle \in E_{\gamma}}\left[\frac{1}{2}\fint dz\rho_{j}(z)\log\left(x-z\right)-\log x\right],\quad i\in V_{\gamma}^{\fullmoon},\nonumber \\
0 & = & -\fint dz\rho_{j}(z)\log\left(x-z\right)+\frac{1}{2}\log\left(\mathtt{q}_{j}\right)+\frac{1}{2}\sum_{f=1}^{w_{j}}\log\left(x^{2}-m_{i,f}^{2}\right)\nonumber \\
 &  & +\frac{\xi_{j}}{2}\log x+\frac{1}{2}\sum_{\left\langle i,j\right\rangle \in E_{\gamma}}\fint dz\rho_{i}(z)\log\left(x-z\right),\quad j\in V_{\gamma}^{\newmoon}.\label{eq:var}
\end{eqnarray}

\subsection{Analytic continuation and the instanton Weyl group}

The limit shape equations (\ref{eq:var}) can be solved in terms of
the amplitude function, which is the generating function of the vacuum
expectation values of all the gauge invariant local observables commuting
with the supercharge $\mathcal{Q}$ \cite{Nekrasov:2012xe}, 
\begin{equation}
\mathcal{Y}_{i}\left(x\right)=\exp\int dz\rho_{i}(z)\log\left(x-z\right).
\end{equation}
We can expand $\mathcal{Y}_{i}\left(x\right)$ as a Laurant series
in $x$,
\begin{equation}
\mathcal{Y}_{i}\left(x\right)=x^{v_{i}}+\sum_{j=-\infty}^{v_{i}-2}\mathcal{Y}_{i,j}x^{j}.
\end{equation}
The function $\mathcal{Y}_{i}\left(x\right)$ is analytic on $\mathbb{C}\setminus\mathscr{C}_{i}$,
and has branch cuts on $\mathscr{C}_{i}$. According to Sokhotsky's
formula, for $x\in\mathscr{C}_{i}$, the principal value
\begin{equation}
\fint dz\rho_{i}(z)\log\left(x-z\right)=\mathcal{Y}_{i}\left(x+\mathtt{i}0\right)\mathcal{Y}_{i}\left(x-\mathtt{i}0\right),
\end{equation}
and the discontinuity across $\mathscr{C}_{i}$ 
\begin{equation}
\frac{\mathcal{Y}_{i}\left(x+\mathtt{i}0\right)}{\mathcal{Y}_{i}\left(x-\mathtt{i}0\right)}=\exp\left(-2\pi\mathtt{i}\int_{-\infty}^{x}dz\rho_{i}(z)\right),
\end{equation}
where $\mathcal{Y}_{i}\left(x\pm\mathtt{i}0\right)$ are the limit
values at the top and the bottom of $\mathscr{I}_{i,\ell}$. If $\mathcal{A}_{i,\ell}$
is a small cycle surrounding the cut $\left[a_{i,\ell}^{-},a_{i,\ell}^{+}\right]$,
then from (\ref{eq:inta}) we know that
\begin{equation}
\pm a_{i,\alpha}=\frac{1}{2\pi\mathtt{i}}\oint_{\mathcal{A}_{i,\pm\alpha}}xd\log\mathcal{Y}_{i}.
\end{equation}
Alternatively, we can view $\mathcal{Y}_{i}(x)$ as a single-valued
holomorphic function living on a Riemann surface, which is the double
cover of the complex plane, glued together along the cuts. 

The limit shape equations (\ref{eq:var}) in terms of $\rho_{i}$
is the same as the nonlinear difference equations on $\mathcal{Y}_{i}\left(x\right)$,
\begin{eqnarray}
\mathcal{Y}_{i}\left(x+\mathtt{i}0\right)\mathcal{Y}_{i}\left(x-\mathtt{i}0\right) & = & \mathtt{q}_{i}x^{4-2\sigma_{i}}\prod_{f=1}^{w_{i}}\left(x^{2}-m_{i,f}^{2}\right)\prod_{\left\langle i,j\right\rangle \in E_{\gamma}}\mathcal{Y}_{j}(x),\quad i\in V_{\gamma}^{\fullmoon},\nonumber \\
\mathcal{Y}_{j}\left(x+\mathtt{i}0\right)\mathcal{Y}_{j}\left(x-\mathtt{i}0\right) & = & \mathtt{q}_{j}x^{\xi_{j}}\prod_{f=1}^{w_{j}}\left(x^{2}-m_{j,f}^{2}\right)\prod_{\left\langle i,j\right\rangle \in E_{\gamma}}\mathcal{Y}_{i}(x),\quad j\in V_{\gamma}^{\newmoon}.\label{eq:Y+-}
\end{eqnarray}
Recall that for $\mathrm{SU}$ quiver gauge theories, the limit shape
equations can be written as \cite{Nekrasov:2012xe}
\begin{equation}
\mathcal{Y}_{i}\left(x+\mathtt{i}0\right)\mathcal{Y}_{i}\left(x-\mathtt{i}0\right)=\mathtt{q}_{i}\prod_{f=1}^{w_{i}}\left(x-m_{i,f}\right)\prod_{e:t(e)=i}\mathcal{Y}_{s(e)}(x+m_{e})\prod_{e:s(e)=i}\mathcal{Y}_{t(e)}(x-m_{e}),\label{eq:YSU}
\end{equation}
where $\mathcal{Y}_{i}(x)$ is the $i$th amplitude function in the
$\mathrm{SU}$ quiver gauge theory, $m_{e}$ is the mass of the bifundamental
hypermultiplet associated with the oriented edge $e$ whose source
and target vertices are $s(e)$ and $t(e)$ respectively. We find
that (\ref{eq:Y+-}) and (\ref{eq:YSU}) are very similar. The differences
arise because we have unoriented bipartite quiver diagrams for $\mathrm{SO}-\mathrm{USp}$
quiver gauge theories and the bifundamental matter fields are half-hypermultiplets
rather than full hypermultiplets.

We can analytically continue $\mathcal{Y}_{i}(x)$ across the cuts
according to (\ref{eq:Y+-}), leading to a multivalued function on
the complex plane. We define the following reflections on a single
vertex,
\begin{eqnarray}
s_{i}:\quad\mathcal{Y}_{i}(x) & \mapsto & x^{4-2\sigma_{i}}P_{i}(x)\mathcal{Y}_{i}(x)^{-1}\prod_{\left\langle i,j\right\rangle \in E_{\gamma}}\mathcal{Y}_{j}(x),\quad i\in V_{\gamma}^{\fullmoon},\nonumber \\
s_{j}:\quad\mathcal{Y}_{j}(x) & \mapsto & P_{j}(x)\mathcal{Y}_{j}(x)^{-1}\prod_{\left\langle i,j\right\rangle \in E_{\gamma}}\mathcal{Y}_{i}(x),\quad j\in V_{\gamma}^{\newmoon}.\label{eq:iWeyl}
\end{eqnarray}
It is easy to check that $s_{i}^{2}=1$ and $s_{i}s_{j}=s_{j}s_{i}$
if $\left\langle i,j\right\rangle \notin E_{\gamma}$. These reflections
generate a group, called the instanton Weyl group $^{i}\mathcal{W}$.
It is the finite Weyl group $\mathcal{W}(\mathfrak{g})$ of the ADE
simple Lie algebra $\mathfrak{g}$ for the Class I theories of type
$\mathfrak{g}$, and is the affine Weyl group $\mathcal{W}(\hat{\mathfrak{g}})$
of the affine Lie algebra $\hat{\mathfrak{g}}$ for the Class II theories
of type $\hat{\mathfrak{g}}$ \cite{Nekrasov:2012xe}. For class III
theories, it needs to be worked out case by case.

The instanton Weyl group $^{i}\mathcal{W}$ is useful due to the following
reason. Notice that although $\mathcal{Y}_{i}(x)$ has discontinuity
across the cut $\mathscr{C}_{i}$, the combination $\mathcal{Y}_{i}(x)+s_{i}\left[\mathcal{Y}_{i}(x)\right]$
is invariant under the reflection $s_{i}$, making it continuous across
the cut $\mathscr{C}_{i}$. There are new discontinuities across other
cuts $\mathscr{C}_{j}$ introduced by $s_{i}\left[\mathcal{Y}_{i}(x)\right]$.
Again these discontinuities can be canceled by acting on other reflections
$s_{j}$. The iteration process will close in a finite or infinite
number of steps and produces an $^{i}\mathcal{W}$-orbit. The resulting
function is manifestly analytic on $\mathbb{C}\setminus\left(\cup_{i}\mathscr{C}_{i}\right)$
and is also continuous across all the cuts due to the $^{i}\mathcal{W}$-invariance.
Therefore, it must be a single-valued analytic function on the whole
complex plane. Our solution to the limit shape equations (\ref{eq:Y+-})
can then be given in terms of a set of $^{i}\mathcal{W}$-invariants.

\subsection{Characters and Seiberg-Witten geometry}

Among all the $^{i}\mathcal{W}$-invariants, we are particularly interested
in the characters $\chi_{i}\left(\mathcal{Y}(x)\right)$ of the $^{i}\mathcal{W}$-orbits
containing $\mathcal{Y}_{i}(x)$,
\begin{equation}
\chi_{i}\left(\mathcal{Y}(x)\right)=\mathcal{Y}_{i}(x)+\cdots=\mathrm{Tr}L_{i}(x),\quad i\in V_{\gamma},
\end{equation}
where $L_{i}(x)$ is a diagonal matrix with entries the components
of $\chi_{i}$,
\begin{equation}
L_{i}(x)=\mathrm{diag}\left\{ L_{i,1},L_{i,2},\cdots\right\} ,
\end{equation}
which is a finite matrix for class I theories, and is an infinite
matrix for class II theories. Each term in $\cdots$ are Laurent polynomials
in $\mathcal{Y}_{j}(x)$ and $\mathcal{Y}_{i}(x)^{-1}$ and the asymptotic
power of $x$ is the same as $\mathcal{Y}_{i}(x)$ near $x=\infty$.
In the weakly coupled limit $\underline{\mathtt{q}}\to0$, $\chi_{i}(x)\to\mathcal{Y}_{i}(x)$.
According to the asymptotic behavior near $x=\infty$, we know that
they must be polynomials in $x$,
\begin{equation}
\chi_{i}\left(\mathcal{Y}(x)\right)=T_{i}(x),\label{eq:Ti}
\end{equation}
where the coefficients of $T_{i}(x)$ are functions of the couplings
$\underline{\mathtt{q}}$, the masses $\underline{m}$, and the coordinate
$\underline{u}$ on the Coulomb branch $\mathcal{B}$. 

The Seiberg-Witten curve of the theory can be uniformly written as
\begin{equation}
\Sigma_{u}:\quad\left.\det\left(1-t^{-1}\zeta(x)L_{i}(x)\right)\right|_{\chi_{j}=T_{j}}=0,\label{eq:spectral}
\end{equation}
where $\zeta(x)$ is a normalization factor to be determined, and
the meromorphic differential takes the canonical form
\begin{equation}
\lambda=x\frac{dt}{t}.
\end{equation}

\section{Seiberg-Witten geometry of linear quiver gauge theories \label{sec:Examples}}

In this section, we shall carefully describe the Seiberg-Witten geometry
of linear quiver gauge theories as an illustrative example.

\subsection{Linear quiver gauge theories}

We consider the class I theory of $A_{r}$-type. The set of vertices
is $V_{\gamma}=\left\{ 1,2,\dots,r\right\} $, and the edges connect
vertices of nearest neighbors. The total gauge group has the structure
\cite{Tachikawa:2009rb,Bhardwaj:2013qia}
\begin{equation}
G=\prod_{i=1}^{r}G_{i}=\cdots\times\mathrm{SO}\left(v_{i}\right)\times\mathrm{USp}\left(v_{i+1}-2\right)\times\cdots,
\end{equation}
where
\begin{equation}
v_{1}<v_{2}<\cdots<v_{\heartsuit-1}<v_{\heartsuit}=\cdots=v_{\spadesuit}>v_{\spadesuit+1}>\cdots v_{r}.
\end{equation}
We refer to the parts to the left of $v_{\heartsuit}$ and to the
right of $v_{\spadesuit}$ as the two tails of the quiver. We also
define $v_{0}=v_{r+1}=0$. The condition (\ref{eq:ADEquiver}) becomes
\begin{equation}
2v_{i}-v_{i-1}-v_{i+1}=2w_{i}+\xi_{i}+2\delta_{i,1}^{\fullmoon}+2\delta_{i,r}^{\fullmoon},
\end{equation}
where $\delta_{i,1}^{\fullmoon}=1$ if $i=1\in V_{\gamma}^{\fullmoon}$
and vanishes otherwise, and similarly for $\delta_{i,r}^{\fullmoon}$.
It is convenient to express $v_{i}$ in the following way,
\begin{equation}
v_{i}=\begin{cases}
\sum_{j=1}^{i}d_{j}, & 1\leq i\leq\heartsuit-1\\
2N, & \heartsuit\leq i\le\spadesuit,\\
\sum_{j=i}^{r}d_{j} & \spadesuit+1\leq i\leq r,
\end{cases}
\end{equation}
where 
\begin{equation}
d_{i}=\begin{cases}
v_{i}-v_{i-1}, & 1\leq i\leq\heartsuit\\
0, & \heartsuit+1\leq i\leq\spadesuit-1\\
v_{i}-v_{i+1}, & \spadesuit\leq i\leq r
\end{cases},
\end{equation}
satisfying
\begin{equation}
d_{1}\geq d_{2}\geq\cdots\geq d_{\heartsuit}\geq0,\quad0\leq d_{\spadesuit}\leq d_{\spadesuit+1}\leq\cdots\leq d_{r},
\end{equation}
and 
\begin{equation}
\sum_{i=1}^{\heartsuit}d_{i}=\sum_{i=\heartsuit}^{\spadesuit}d_{i}=v_{\heartsuit}.
\end{equation}
Therefore, it is natural to associate each tail with a Young tableau
\cite{Tachikawa:2009rb}. The Young tableau associated with the left
tail has row lengths being nonincreasing integers $d_{1}-2\delta_{1,1}^{\fullmoon},d_{2},\cdots,d_{\heartsuit}$,
and the difference between the $i$th and the $(i+1)$th row lengths
gives $2w_{i}+\xi_{i}$. We also have a similar Young tableau associated
with the right tail. 

\subsection{Seiberg-Witten curve}

The instanton Weyl group $^{i}\mathcal{W}$ for the quiver of $A_{r}$-type
is the symmetric group $\mathcal{S}_{r+1}$, which is generated by
\begin{equation}
\mathcal{Y}_{i}\mapsto P_{i}\mathcal{Y}_{i}^{-1}\mathcal{Y}_{i-1}\mathcal{Y}_{i+1},\quad i=1,\cdots,r,
\end{equation}
where
\begin{equation}
P_{i}(x)=\mathtt{q}_{i}x^{2\delta_{i,1}^{\fullmoon}+2\delta_{i,r}^{\fullmoon}+\xi_{i}}\prod_{f=1}^{w_{i}}\left(x^{2}-m_{i,f}^{2}\right).
\end{equation}
This is very similar to the case for the $\mathrm{SU}$ linear quiver
gauge theories. Under a chain of Weyl reflections, we get a $^{i}\mathcal{W}$-orbit
starting from $\mathcal{Y}_{1}$,
\begin{equation}
\mathcal{Y}_{1}\xrightarrow{s_{1}}P^{[1]}\mathcal{Y}_{1}^{-1}\mathcal{Y}_{2}\xrightarrow{s_{2}}P^{[2]}\mathcal{Y}_{2}^{-1}\mathcal{Y}_{3}\xrightarrow{s_{3}}\cdots\xrightarrow{s_{r-1}}P^{[r-1]}\mathcal{Y}_{r-1}^{-1}\mathcal{Y}_{r}\xrightarrow{s_{r}}P^{[r]}\mathcal{Y}_{r}^{-1},
\end{equation}
where 
\begin{equation}
P^{[i]}=\prod_{j=1}^{i}P_{j},\quad i=1,\cdots,r.
\end{equation}
The instanton Weyl group $^{i}\mathcal{W}$ acts by permuting the
eigenvalues of the matrix $L_{1}(x)$,
\begin{equation}
L_{1}(x)=\mathrm{diag}\left\{ \mathcal{Y}_{1},P^{[1]}\mathcal{Y}_{1}^{-1}\mathcal{Y}_{2},P^{[2]}\mathcal{Y}_{2}^{-1}\mathcal{Y}_{3},\cdots,P^{[r-1]}\mathcal{Y}_{r-1}^{-1}\mathcal{Y}_{r},P^{[r]}\mathcal{Y}_{r}^{-1}\right\} .
\end{equation}
We can check that the $i$th character is given by 
\begin{equation}
\chi_{i}=\left(P^{[1,i-1]}\right)^{-1}\mathtt{e}_{i}\left(\mathcal{Y}_{1},P^{[1]}\mathcal{Y}_{1}^{-1}\mathcal{Y}_{2},P^{[2]}\mathcal{Y}_{2}^{-1}\mathcal{Y}_{3},\cdots,P^{[r-1]}\mathcal{Y}_{r-1}^{-1}\mathcal{Y}_{r},P^{[r]}\mathcal{Y}_{r}^{-1}\right),\label{eq:character}
\end{equation}
where we introduce the notation
\begin{equation}
P^{[i,j]}=\prod_{n=i}^{j}P^{[n]},
\end{equation}
and $\mathtt{e}_{i}$ is the $i$th elementary symmetric polynomial
\begin{equation}
\mathtt{e}_{0}\left(x_{1},\cdots,x_{n}\right)=1,\quad\mathtt{e}_{i}\left(x_{1},\cdots,x_{n}\right)=\sum_{1\leq j_{1}<\cdots<j_{i}\leq n}x_{j_{1}}\cdots x_{j_{i}},\quad1\leq i\leq n.
\end{equation}
In fact, the elementary symmetric polynomials are totally symmetric
in the variables, and $\mathcal{Y}_{i}$ is one term in $\chi_{i}$
(\ref{eq:character}),
\begin{equation}
\left(P^{[1,i-1]}\right)^{-1}\left(\mathcal{Y}_{1}\right)\left(P^{[1]}\mathcal{Y}_{1}^{-1}\mathcal{Y}_{2}\right)\cdots\left(P^{[i-1]}\mathcal{Y}_{i-1}^{-1}\mathcal{Y}_{i}\right)=\mathcal{Y}_{i}.
\end{equation}
Therefore, (\ref{eq:character}) is the characters of the $^{i}\mathcal{W}$-orbits
starting from $\mathcal{Y}_{i}$ for $i=1,\cdots,r$. We also define
\begin{equation}
\chi_{0}=\chi_{r+1}=1.
\end{equation}

We expand the character polynomial in terms of the elementary symmetric
polynomials 
\begin{eqnarray}
\det\left(t-\zeta(x)L_{1}(x)\right) & = & \sum_{i=0}^{r+1}(-1)^{i}\zeta(x)^{i}\mathtt{e}_{i}\left(L_{1}(x)\right)t^{r+1-i}\nonumber \\
 & = & \sum_{i=0}^{r+1}(-1)^{i}\zeta(x)^{i}P^{[1,i-1]}\chi_{i}(x)t^{r+1-i}.
\end{eqnarray}
Substituting $\chi_{i}(x)$ by $T_{i}(x)$, we get explicitly the
Seiberg-Witten curve,
\begin{equation}
\Sigma_{u}:\quad t^{r+1}+\sum_{i=1}^{r}(-1)^{i}\zeta(x)^{i}P^{[1,i-1]}T_{i}(x)t^{r+1-i}+(-1)^{r}\zeta(x)^{r}P^{[1,r]}=0,\label{eq:SWcurve}
\end{equation}
where the polynomial $T_{i}(x)$ takes the form
\begin{equation}
T_{i}(x)=x^{v_{i}}\left(T_{i,0}+T_{i,1}x^{-2}+T_{i,2}x^{-4}+\cdots+T_{i,\left\lfloor \frac{v_{i}}{2}\right\rfloor }x^{-2\left\lfloor \frac{v_{i}}{2}\right\rfloor }\right),\quad i=1,\cdots,r,
\end{equation}
with $\left\lfloor \alpha\right\rfloor $ being the greatest integer
less than or equal to $\alpha$. Notice that only even powers of $x$
can appear in the bracket. The leading coefficient depends only on
the couplings
\begin{equation}
T_{i,0}=\left(\prod_{j=1}^{i-1}\mathtt{q}_{j}^{j-i}\right)e_{i}\left(1,\mathtt{q}_{1},\mathtt{q}_{1}\mathtt{q}_{2},\cdots,\mathtt{q}_{1}\mathtt{q}_{2}\cdots\mathtt{q}_{r}\right).
\end{equation}
The next-to-leading coefficient $T_{i,1}$ is a function of the couplings
and the masses. The remaining coefficients encode the information
of vacuum expectation values of Coulomb branch operators. The Seiberg-Witten
curves (\ref{eq:SWcurve}) with different normalization factor $\zeta(x)$
contain the same physical information and are related to each other
by a change of variables $(t,x)$. The Seiberg-Witten curve obtained
in \cite{Landsteiner:1997vd,Brandhuber:1997cc} by lifting a system
of D4/NS5/D6-branes with orientifolds in type IIA string theory to
M-theory matches (\ref{eq:SWcurve}) with $\zeta(x)=-1$.

\section{Further developments \label{sec:Further}}

In this paper, we derive the Seiberg-Witten geometry of $\mathrm{SO}-\mathrm{USp}$
quiver gauge theories using the instanton counting method, with an
emphasize on linear quiver gauge theories. Our discussion can be straightforwardly
lifted to five-dimensional $\mathcal{N}=1$ theories compactified
on $S^{1}$ or six-dimensional $\mathcal{N}=\left(1,0\right)$ theories
compactified on $T^{2}$. For the partition function, the equivariant
cohomology should be replaced by corresponding K-theoretical or elliptic
version. The corresponding Seiberg-Witten geometry can be derived
in the same way. 

After solving the linear quiver gauge theories, it is very natural
to also work out the other quiver gauge theories. Indeed, if the quiver
is one of the ADE or affine ADE Dynkin diagrams, the analysis would
be very similar to the corresponding $\mathrm{SU}$ quiver gauge theories
\cite{Nekrasov:2012xe}. However, it is more interesting to consider
the non-Dynkin type quivers. Even the Seiberg-Witten solutions to
most of them are unknown so far. The instanton counting method seems
to be the most promising approach to solve them. We will discuss all
these cases in Part II of our article.

There are many other open questions that will be studied in the future.
We can study the Bethe/gauge correspondence between the supersymmetric
gauge theories and quantum integrable systems by sending only $\varepsilon_{2}\to0$
while keeping $\varepsilon_{1}=\hbar$ finite \cite{Nekrasov:2009rc},
generalizing the derivation for $\mathrm{SU}$ quiver gauge theories
in \cite{Fucito:2012xc,Nekrasov:2013xda}. The effective twisted superpotential
can be obtained from the partition function via
\begin{equation}
\widetilde{W}^{\mathrm{eff}}\left(\underline{\mathtt{q}};\underline{a},\underline{m};\hbar\right)=-\lim_{\varepsilon_{2}\to0}\varepsilon_{2}\log\mathcal{Z}\left(\underline{\mathtt{q}};\underline{a},\underline{m};\varepsilon_{1}=\hbar,\varepsilon_{2}\right)+\widetilde{W}^{\infty}\left(\underline{a},\underline{m};\hbar\right),
\end{equation}
where $\widetilde{W}^{\infty}$ is the possible perturbative contribution
from the boundary conditions at infinity, and $\widetilde{W}^{\mathrm{eff}}$
is identified with the Yang-Yang function of some quantum integrable
system \cite{Nekrasov:2009uh,Nekrasov:2009ui,Nekrasov:2009rc,Nekrasov:2010ka,Nekrasov:2011bc,Jeong:2018qpc}. 

We can go one step further and consider the situation where both $\varepsilon_{1}$
and $\varepsilon_{2}$ are finite. In this case we shall introduce
an interesting class of gauge-invariant observables, $\mathscr{Y}_{i}(x)$,
whose vacuum expectation values are $\mathcal{Y}_{i}(x)$. We should
be able to define the so-called qq-characters $\mathscr{X}_{i}(x)$,
which are composite operators built from $\mathscr{Y}_{i}(x)$ and
satisfy the nonperturbative Dyson-Schwinger equations \cite{Nekrasov:2015wsu,Nekrasov:2016qym}.
The theory of qq-characters play an important role in the study of
$\mathrm{SU}$ quiver gauge theories. For example, they can be used
to derive the Belavin-Polyakov-Zamolodchikov equations from the field
theory point of view \cite{Nekrasov:2017gzb,Jeong:2017mfh}. It is
also interesting to study the relations among the vacuum expectation
values of chiral operators in the $\Omega$-background \cite{Jeong:2019fgx}.
A closely related issue is the study of the gauge origami \cite{Nekrasov:2016ydq}
in the presence of the orientifold plane.

In recent years much of the investigation of supersymmetric gauge
theories has involved the presence of nonlocal operators. We can naturally
study surface operators in the $\Omega$-background \cite{Alday:2009fs,Alday:2010vg,Awata:2010bz,Kanno:2011fw,Frenkel:2015rda,Gomis:2016ljm,Pan:2016fbl,Nekrasov:2017rqy,Jeong:2017pai}.
Unfortunately, almost nothing has been said about surface operators
in the $\Omega$-background when the gauge group is $\mathrm{SO}/\mathrm{USp}$.

Finally, we would like to emphasize that the analysis of $\mathrm{SO}-\mathrm{USp}$
quiver gauge theories is not as rigorous as that of $\mathrm{SU}$
quiver gauge theories. The hazards come not only from the treatment
of the half-hypermultiplets, but also the noncompactness of the moduli
space of $\mathrm{SO}/\mathrm{USp}$ instantons. When we take the
flat space limit, we neglect a lot of information of the partition
function in the $\Omega$-background, and many potentially problematic
issues are avoided. Hence we cannot say that we fully understand the
$\Omega$-background before we have completed the above generalizations
from $\mathrm{SU}$ to $\mathrm{SO}-\mathrm{USp}$ quiver gauge theories.


\acknowledgments
The work was supported by the U.S. Department of Energy under Grant No. DE-SC0010008. 
The author would like to thank Saebyeok Jeong, Nikita Nekrasov, and Wenbin Yan for helpful discussions. 
The author also gratefully acknowledges support from the Simons Center for Geometry and Physics, Stony Brook University at which part of the research for this paper was performed.


\appendix

\section{Instanton moduli space \label{sec:Instanton}}

In this appendix, we review some properties of the moduli spaces $\mathfrak{M}_{G,k}$
of framed $G$-instantons on $\mathbb{C}^{2}$ with instanton charge
$k$,
\begin{equation}
\mathfrak{M}_{G,k}=\left.\left\{ A\in\mathcal{A}_{G}\Big|F+\ast F=0,k=\frac{1}{16\pi^{2}h^{\vee}}\int_{\mathbb{C}^{2}}\mathrm{Tr}_{\mathrm{adj}}F\wedge F\right\} \right/\mathcal{G}_{\infty},
\end{equation}
where $\mathcal{A}_{G}$ is the space of $G$-connections, $h^{\vee}$
is the dual Coxeter number for the Lie algebra of $G$, $\mathrm{Tr}_{\mathrm{adj}}$
is the trace in the adjoint representation, and $\mathcal{G}_{\infty}$
is the group of gauge transformations that are identity at infinity.
For $G$ being a classical group, Atiyah, Drinfeld, Hitchin and Manin
(ADHM) found a description of $\mathfrak{M}_{G,k}$ in terms of solutions
to quadratic equations for certain finite-dimensional matrices \cite{Atiyah:1978ri}. 

It is useful to introduce the following notations. Let $S^{\pm}$
be the positive and negative spin bundles, and the line bundle $L=\mathcal{K}_{\mathbb{C}^{2}}^{1/2}$
be the half canonical bundle of $\mathbb{C}^{2}$. The group of rotations
on $\mathbb{C}^{2}$ with a fixed translationally invariant symplectic
form is $G_{R}=\mathrm{U}(2)\simeq\mathrm{SU}(2)_{-}\times\mathrm{U}(1)_{+}\subset\mathrm{Spin}(4)$,
under which $S^{+}$ splits as $L\oplus L^{-1}$.

\subsection{$\mathrm{U}(n)$ instantons}

We start with the moduli space of $\mathrm{U}(n)$ instantons. We
introduce a quartet of linear operators
\begin{equation}
\left(B_{1},B_{2},I,J\right)\in\mathrm{Hom}(\mathbf{K},\mathbf{K})\otimes\mathbb{C}^{2}\bigoplus\mathrm{Hom}(\mathbf{N},\mathbf{K})\bigoplus\mathrm{Hom}(\mathbf{K},\mathbf{N}),
\end{equation}
where $\mathbf{K}$ and $\mathbf{N}$ are two complex vector spaces
of dimension $k$ and $n$, respectively. The moduli space of framed
$\mathrm{U}(n)$ instantons on $\mathbb{C}^{2}$ with instanton charge
$k$ is given by the regular locus of the hyperkahler quotient of
the space of operators $\left(B_{1},B_{2},I,J\right)$ by the $\mathrm{U}(k)$
action, 
\begin{equation}
\mathfrak{M}_{\mathrm{U}(n),k}\cong\left\{ \left(B_{1},B_{2},I,J\right)\big|\mu_{\mathbb{C}}=0,\mu_{\mathbb{R}}=0\right\} ^{\mathrm{reg}}/\mathrm{U}(k),
\end{equation}
where the ADHM moment maps are
\begin{eqnarray}
\mu_{\mathbb{C}} & = & \left[B_{1},B_{2}\right]+IJ,\\
\mu_{\mathbb{R}} & = & \left[B_{1},B_{1}^{\dagger}\right]+\left[B_{2},B_{2}^{\dagger}\right]+II^{\dagger}-J^{\dagger}J,
\end{eqnarray}
the action of $g\in\mathrm{U}(k)$ is 
\begin{equation}
g\cdot\left(B_{1},B_{2},I,J\right)=\left(gB_{1}g^{-1},gB_{2}g^{-1},gI,Jg^{-1}\right),\quad g\in\mathrm{U}(k),
\end{equation}
and the regularity requires that the group action of $\mathrm{U}(k)$
is free on the solution $\left(B_{1},B_{2},I,J\right)$. 

The ADHM construction can be represented by the following complex,
\begin{equation}
0\to\mathbf{K}\otimes\mathcal{L}^{-1}\xrightarrow{\alpha}\mathbf{K}\otimes\mathcal{S}^{-}\bigoplus\mathbf{N}\xrightarrow{\beta}\mathbf{K}\otimes\mathcal{L}\to0,\label{eq:E}
\end{equation}
where $\mathcal{L}$ and $\mathcal{S}^{-}$ are the fibers of $L$
and $S^{-}$, respectively, and
\begin{equation}
\alpha=\begin{pmatrix}B_{1}-z_{1}\\
B_{2}-z_{2}\\
J
\end{pmatrix},\quad\beta=\begin{pmatrix}-B_{2}+z_{2}, & B_{1}-z_{1}, & I\end{pmatrix}.
\end{equation}
From the middle cohomology of the complex (\ref{eq:E}), we can form
the virtual universal bundle $\mathcal{E}$ on $\mathbb{C}^{2}\times\mathfrak{M}_{\mathrm{U}(n),k}$
by
\begin{equation}
\mathcal{E}=\mathbf{N}\bigoplus\mathbf{K}\otimes\left(S^{-}\ominus S^{+}\right).\label{eq:ENK}
\end{equation}

The moduli space $\mathfrak{M}_{\mathrm{U}(n),k}$ has singularities
due to pointlike instantons. In order to make the localization computations
appropriate, we may work with another space
\begin{equation}
\mathfrak{M}_{\mathrm{U}(n),k}^{\zeta}\cong\left\{ \left(B_{1},B_{2},I,J\right)\big|\mu_{\mathbb{C}}=0,\mu_{\mathbb{R}}=\zeta\cdot\mathbb{I}_{\mathbf{K}}\right\} /\mathrm{U}(k),
\end{equation}
where $\zeta>0$ is a constant. The moduli space $\mathfrak{M}_{\mathrm{U}(n),k}^{\zeta}$
is a $4nk$ dimensional smooth manifold, with the metric inherited
from the flat metric on $\left(B_{1},B_{2},I,J\right)$, and we can
again form the universal sheaf $\mathcal{E}$ on $\mathbb{C}^{2}\times\mathfrak{M}_{\mathrm{U}(n),k}$
similar to (\ref{eq:ENK}). An equivalent description of $\mathfrak{M}_{\mathrm{U}(n),k}^{\zeta}$
can be given as
\begin{equation}
\mathfrak{M}_{\mathrm{U}(n),k}^{\zeta}\cong\left.\left\{ \left(B_{1},B_{2},I,J\right)\big|\mu_{\mathbb{C}}=0,\mathbb{C}\left[B_{1},B_{2}\right]I(\mathbf{N})=\mathbf{K}\right\} \right/\mathrm{GL}(\mathbf{K}).
\end{equation}
It was shown in \cite{Nekrasov:1998ss} that $\mathfrak{M}_{\mathrm{U}(n),k}^{\zeta}$
describes the moduli space of framed $\mathrm{U}(n)$ instantons on
noncommutative $\mathbb{C}^{2}$ with instanton charge $k$. Mathematically,
$\mathfrak{M}_{\mathrm{U}(n),k}^{\zeta}$ is the moduli space of framed
torsion free sheaves $\left(E,\Phi:E|_{\mathbb{CP}_{\infty}^{1}}\xrightarrow{\sim}\mathcal{O}_{\mathbb{CP}_{\infty}^{1}}^{\oplus n}\right)$
of rank $N$ on $\mathbb{CP}^{2}=\mathbb{C}\cup\mathbb{CP}_{\infty}^{1}$,
with $\left\langle \mathrm{\mathrm{ch}}_{2}(E),\left[\mathbb{CP}^{2}\right]\right\rangle =k$
\cite{nakajima1999lectures}. 

There is a natural $\mathrm{GL}(\mathbf{N})$ action acting on the
moduli space,
\begin{equation}
\rho\cdot\left(B_{1},B_{2},I,J\right)=\left(B_{1},B_{2},I\rho^{-1},\rho J\right),\quad\rho\in\mathrm{GL}(\mathbf{N}).\label{eq:GLW}
\end{equation}
The central $\mathrm{GL}\left(1,\mathbb{C}\right)$ subgroup acts
trivially due to the equivalence under $\mathrm{GL}\left(1,\mathbb{C}\right)\subset\mathrm{GL}(\mathbf{K})$.
Meanwhile, the rotation symmetry of $\mathbb{C}^{2}$ induces a $\left(\mathbb{C}^{*}\right)^{2}$-action
on the moduli space via

\begin{equation}
\left(B_{1},B_{2},I,J\right)\mapsto\left(q_{1}B_{1},q_{2}B_{2},I,q_{1}q_{2}J\right),\quad q_{1},q_{2}\in\mathbb{C}^{*}.\label{eq:T2}
\end{equation}

\subsection{$\mathrm{SO}/\mathrm{USp}$ instantons}

The ADHM construction for $\mathrm{SO}/\mathrm{USp}$ instantons can
be obtained by a projection of the $\mathrm{U}(n)$ instantons. Here
we follow the description given in \cite{BRYAN2000331}. We define
$\mathrm{SO}(n)$ to be the special unitary transformations on $\mathbb{C}^{n}$
that preserve its real structure $\Phi_{r}$, and define $\mathrm{USp}(2n)$
to be the special unitary transformations on $\mathbb{C}^{2n}$ that
preserve its symplectic structure $\Phi_{s}$.

For $\mathrm{SO}(n)$, we consider linear operators
\begin{equation}
\left(B_{1},B_{2},J\right)\in\mathrm{Hom}(\mathbf{K},\mathbf{K})\otimes\mathbb{C}^{2}\bigoplus\mathrm{Hom}(\mathbf{K},\mathbf{N}),
\end{equation}
where $\mathbf{K}$ and $\mathbf{N}$ are two complex vector spaces
of dimension $2k$ and $n$, respectively, together with a symplectic
structure $\Phi_{s}$ on $\mathbf{K}$ and a real structure $\Phi_{r}$
on $\mathbf{N}$. The moduli space of framed $\mathrm{SO}(n)$ instantons
is given by
\begin{equation}
\mathfrak{M}_{\mathrm{SO}(n),k}=\left.\left\{ \left(B_{1},B_{2},J\right)\big|\Phi_{s}B_{1},\Phi_{s}B_{2}\in\wedge^{2}\mathbf{K}^{*},\Phi_{s}\left[B_{1},B_{2}\right]-J^{*}\Phi_{r}J=0\right\} ^{\mathrm{reg}}\right/\mathrm{USp}(2k).
\end{equation}
Similarly, for $\mathrm{USp}(2n)$, we consider linear operators
\begin{equation}
\left(B_{1},B_{2},J\right)\in\mathrm{Hom}(\mathbf{K},\mathbf{K})\otimes\mathbb{C}^{2}\bigoplus\mathrm{Hom}(\mathbf{K},\mathbf{N}),
\end{equation}
where $\mathbf{K}$ and $\mathbf{N}$ are two complex vector spaces
of dimension $k$ and $2n$, respectively, together with a real structure
$\Phi_{r}$ on $\mathbf{K}$ and a symplectic structure $\Phi_{s}$
on $\mathbf{N}$. The moduli space of framed $\mathrm{USp}(2n)$ instantons
is given by
\begin{equation}
\mathfrak{M}_{\mathrm{USp}(2n),k}=\left.\left\{ \left(B_{1},B_{2},J\right)\big|\Phi_{r}B_{1},\Phi_{r}B_{2}\in S^{2}\mathbf{K}^{*},\Phi_{r}\left[B_{1},B_{2}\right]-J^{*}\Phi_{s}J=0\right\} ^{\mathrm{reg}}\right/\mathrm{O}(k).
\end{equation}

For both $\mathrm{SO}(n)$ and $\mathrm{USp}(2n)$ instantons, we
do not know the compactification of the moduli space of framed instantons
which admits a universal bundle with the universal instanton connection
over $\mathfrak{M}_{G,k}\times\mathbb{C}^{2}$. Nevertheless, we still
have the ADHM complex 
\begin{equation}
0\to\mathbf{K}\otimes\mathcal{L}^{-1}\xrightarrow{\alpha}\mathbf{K}\otimes\mathcal{S}^{-}\bigoplus\mathbf{N}\xrightarrow{\alpha^{*}\beta^{*}}\mathbf{K}^{*}\otimes\mathcal{L}\to0,
\end{equation}
where 
\begin{equation}
\alpha=\begin{pmatrix}B_{1}-z_{1}\\
B_{2}-z_{2}\\
J
\end{pmatrix},\quad\beta^{\mathrm{SO}}=\begin{pmatrix}0 & \Phi_{s} & 0\\
-\Phi_{s} & 0 & 0\\
0 & 0 & -\Phi_{r}
\end{pmatrix},\quad\beta^{\mathrm{Sp}}=\begin{pmatrix}0 & \Phi_{r} & 0\\
-\Phi_{r} & 0 & 0\\
0 & 0 & -\Phi_{s}
\end{pmatrix}.
\end{equation}
The induced $\left(\mathbb{C}^{*}\right)^{2}$-action on the moduli
space is given by

\begin{equation}
\left(B_{1},B_{2},J\right)\mapsto\left(q_{1}B_{1},q_{2}B_{2},q_{+}J\right).\label{eq:T2SOSp}
\end{equation}


\bibliographystyle{JHEP}
\bibliography{Ref}

\end{document}